\documentclass[twocolumn,tighten]{aastex63}

\usepackage{subfigure}
\usepackage{booktabs}
\usepackage{amsmath}
\usepackage{soul}

\newcommand{\hi}{\ion{H}{1}}
\newcommand{\hin}{\ion{H}{1}}

\def\hi{{{\rm H}\,{\sc i}~}}
\def\hin{{{\rm H}\,{\sc i}}}

\shorttitle{diffuse cold CGM}
\shortauthors{Das et al.}
\graphicspath{{./}{figures/}}

\begin{document}

\title{Detection of the diffuse \hi emission in the Circumgalactic Medium of NGC\,891 and NGC\,4565}

\correspondingauthor{Sanskriti Das}
\email{das.244@buckeyemail.osu.edu}

\author[0000-0002-9069-7061]{Sanskriti Das}
\affiliation{Department of Astronomy, The Ohio State University, 140 West 18th Avenue, Columbus, OH 43210, USA}

\author{Amy Sardone}
\affiliation{Department of Astronomy, The Ohio State University, 140 West 18th Avenue, Columbus, OH 43210, USA}
\affil{Center for Cosmology and Astroparticle Physics, 191 West Woodruff Avenue, Columbus, OH 43210, USA}
\author{Adam K. Leroy}
\affiliation{Department of Astronomy, The Ohio State University, 140 West 18th Avenue, Columbus, OH 43210, USA}
\affil{Center for Cosmology and Astroparticle Physics, 191 West Woodruff Avenue, Columbus, OH 43210, USA}
\author{Smita Mathur}
\affiliation{Department of Astronomy, The Ohio State University, 140 West 18th Avenue, Columbus, OH 43210, USA}
\affil{Center for Cosmology and Astroparticle Physics, 191 West Woodruff Avenue, Columbus, OH 43210, USA}
\author{Molly Gallagher}
\affil{Department of Physics,University of Connecticut, 2152 Hillside Road, Storrs, CT 06268, USA}
\author{Nickolas M. Pingel}
\affil{Research Scool of Astronomy \& Astrophysics, Australian National University, Canberra, ACT 2611, Australia}
\author{D. J. Pisano}
\affil{Department of Physics and Astronomy, West Virginia University, PO Box
6315, Morgantown, WV 26506, USA}
\affil{Center for Gravitational Waves and Cosmology, West Virginia University, Chestnut Ridge Research Building, Morgantown, WV 26505, USA}
\affil{Adjunct Astronomer at Green Bank Observatory, Green Bank, WV, USA}
\author{George Heald}
\affil{CSIRO Astronomy and Space Science, PO Box 1130, Bentley, WA 6102, Australia}


\begin{abstract}
\noindent We present detections of 21-cm emission from neutral hydrogen (\hin) in the circumgalactic medium (CGM) of the local edge-on galaxies NGC\,891 and NGC\,4565 using the Robert C. Byrd Green Bank Telescope (GBT). With our 5$\sigma$ sensitivity of $8.2 \times 10^{16}$ cm$^{-2}$ calculated over a 20 km s$^{-1}$ channel, we achieve $>5\sigma$ detections out to $90-120$\,kpc along the minor axes. 
The velocity width of the CGM emission is as large as that of the disk $\approx 500$ km s$^{-1}$, indicating the existence of a diffuse component permeating the halo. We compare our GBT measurements with interferometric data from the Westerbork Synthesis Radio Telescope (WSRT). The WSRT maps the \hi emission from the disk at high S/N but has limited surface brightness sensitivity at the angular scales probed with the GBT. After convolving the WSRT data to the spatial resolution of the GBT (FWHM = 9.1$'$), we find that the emission detected by the WSRT accounts for $48^{+15}_{-25}$\%({$58^{+4}_{-18}$\%}) of the total flux recovered by the GBT from the CGM of NGC\,891(NGC\,4565). The existence of significant GBT-only flux suggests the presence of a large amount of diffuse, low column density \hi emission in the CGM. For reasonable assumptions, the extended diffuse \hi could account for $5.2\pm0.9$\% and {$2.0\pm0.8$\%} of the total \hi emission of NGC\,891 and NGC\,4565.
\end{abstract}

\keywords{21-cm emission --- neutral CGM --- GBT---galaxies: individual (NGC\,891, NGC\,4565)}


\section{Introduction} \label{sec:intro}
\begin{figure*}
        \centering
        \includegraphics[trim = 5 5 5 0, clip,scale=0.65]{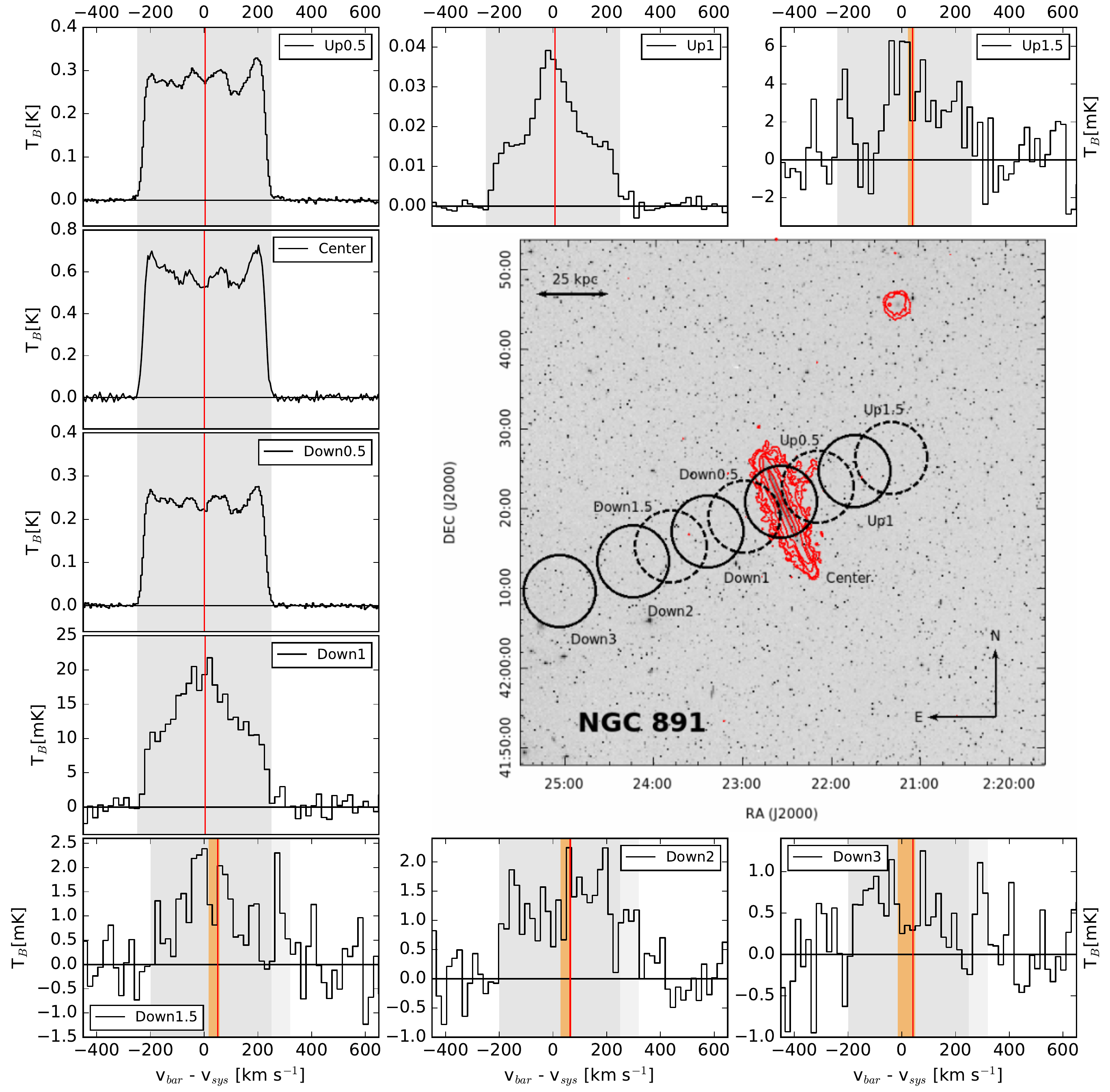}
        \caption{The 21-cm spectra for the observed GBT pointings along the minor axis of NGC\,891, smoothed and re-binned at 20 km s$^{-1}$. The pointings are illustrated on the Digitized Sky Survey (DSS) image of the galaxy. The nomenclature of the pointing is based on its position along the minor axis of the galaxy, see Table \ref{tab:obs} for more details. The red contours show the integrated 21-cm intensity from the HALOGAS survey (adopted from \cite{Oosterloo2007}), with the lowest contour showing a column density of 1$\times$10$^{19}$ cm$^{-2}$. We calculate the integrated intensity at each pointing by summing the brightness temperature over the velocity range shaded in gray. At the pointings far from the disk, we detect emission signature at a level of N(\hin) = $4.8\times10^{17}-2.6\times10^{18}$ cm$^{-2}$, with 6--9$\sigma$ significance (including systematic uncertainties). The average velocities and the associated uncertainties are plotted in red vertical lines and the orange areas. The spectra, as described in \S \ref{sec:obsred}, from the top-right to the bottom-right of the plot in counter-clockwise order are obtained from the rightmost pointing to the leftmost pointing in the DSS image along the minor axis, respectively.}
        \label{fig:NGC891}
\end{figure*}
\begin{figure*}
        \centering
        \includegraphics[trim = 5 0 5 0, clip,scale=0.65]{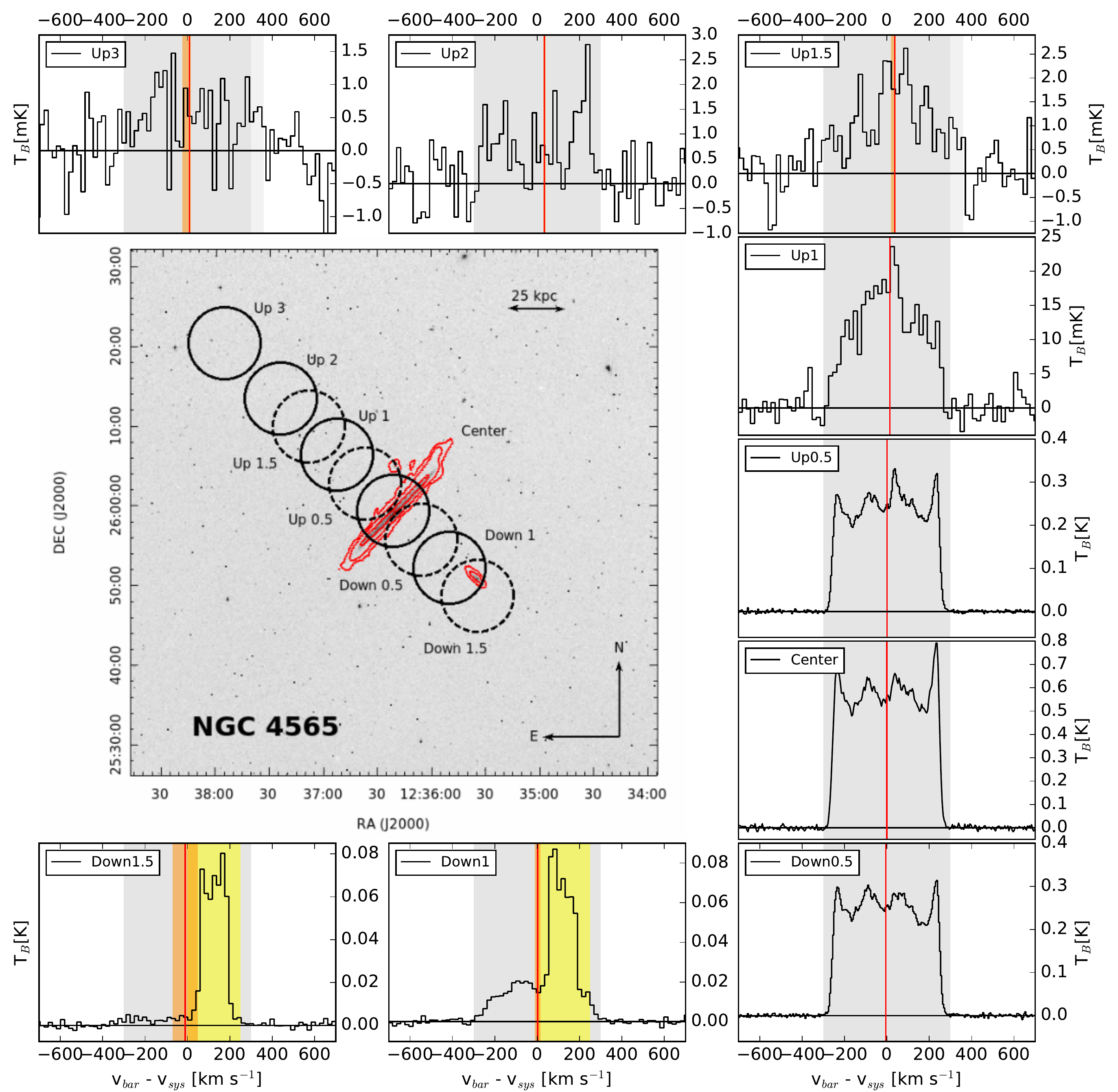}
    \caption{The 21-cm spectra for the observed GBT pointings along the minor axis of NGC\,4565, smoothed and re-binned at 20 km s$^{-1}$. The pointings are illustrated on the Digitized Sky Survey (DSS) image of the galaxy. The nomenclature of the pointing is based on its position along the minor axis of the galaxy, see Table \ref{tab:obs} for more details. The red contours show the integrated 21-cm intensity from the HALOGAS survey (adopted from \cite{Oosterloo2007}), with the lowest contour showing a column density of 1$\times$10$^{19}$ cm$^{-2}$. We calculate the integrated intensity at each pointing by summing the brightness temperature over the velocity range shaded in gray. At the pointings far from the disk, we detect emission signature at a level of N(\hin) = $6.5\times10^{17}-2.7\times10^{18}$ cm$^{-2}$, with 7--11$\sigma$ significance (including systematic uncertainties). The average velocities and the associated uncertainties are plotted in red vertical lines and the orange areas. The spectra, as described in \S \ref{sec:obsred}, from the top-left to the bottom-left of the plot in clockwise order are obtained from the leftmost pointing to the rightmost pointing in the DSS image along the minor axis, respectively. The areas shaded in yellow are dominated by the disk emission from a companion galaxy NGC\,4562, southwest to NGC\,4565. We exclude this velocity range from our analysis, interpolating across the blanked region.}
    \label{fig:NGC4565}
\end{figure*}
\noindent The circumgalactic medium (CGM) is the halo of multi-phase gas and dust outside the stellar disks of galaxies, but within their virial radii \citep{Tumlinson2017}. 
Numerical simulations predict that precipitation of the condensed hot CGM and filamentary ``cold" mode accretion from the intergalactic medium (IGM) are the likely mechanisms to fuel star formation \citep{Keres2005,Joung2012,Nelson2013,Voit2017,Fraternali2017}. ``Cold" in this context refers to the gas that has not been heated to the virial temperature. Despite the ``cold" label, this gas is still mostly ionized by the UV and X-ray background. Nevertheless, studying the abundance of neutral hydrogen (\hin) in the CGM can be extremely useful. {\hi in the CGM can trace the tidal interaction, ram-pressure stripping and mergers between galaxies \citep{Hibbard1996,Wolfe2013,Odekon2016}}, and the gas {accreting/infalling toward the disk, outflowing from the galactic nucleus to the halo \citep[see the reviews by][for more details]{Sancisi2008,Putman2012,Veilleux2020}}. 

\hi around external galaxies at large impact parameter has so far mostly been studied using UV absorption lines \citep[e.g.,][]{Tripp2008,Stocke2010,Ribaudo2011,Tilton2012,Tumlinson2013,Wotta2016,Prochaska2017}. Absorption studies of the Milky Way halo show the evidence for a large amount of atomic gas at N(\hin) $\geq 10^{18}$ cm$^{-2}$ \citep[and references therein]{Richter2017}. Absorption studies are very useful to accurately probe the column density of \hi along the line of sight, but they require a bright background source. As a result, absorption measurements reveal the average spatial structure of the CGM only after combining many quasar-galaxy pairs. 
On the other hand, emission studies do not rely on background sources. Therefore, they can constrain the large-scale structure of neutral gas in individual galaxy halos in an unbiased, complete way.

\hi emission outside spiral galaxies has been observed for many years. Over the past decade, there have been several large interferometric programs to map the circumgalactic 21-cm emission around nearby spirals, including Hydrogen Accretion in LOcal GAlaxies Survey \citep[HALOGAS,][]{Heald2011}\footnote{\url{https://www.astron.nl/halogas/data.php}}, and The Local Volume \hi Survey \citep[LVHIS,][]{Koribalski2018}. These observations have revealed a great detail about the morphology, size, kinematics and the mass of the extended \hi around galaxies, including extended filaments and the presence of extragalactic analogs to the Milky Way's intermediate- and high-velocity clouds.

The point source sensitivity of the interferometric surveys mentioned above is very high, and they typically achieve $\sim 1'$ angular resolution. However, {due to the finite size of the dishes, there is a minimum possible spacing (i.e., the diameter of each dish) between neighboring telescopes of an interferometric array. This results in a gap in the  \textit{uv}-coverage, which is called the ``short-spacing" problem. It limits the sensitivity of the interferometers to low surface brightness and large angular scales (e.g., the smallest baseline length of WSRT is 36\,m, which translates to a maximum
recoverable angular scale of $20'$).} As any truly ``diffuse" component of the neutral CGM will tend to have a very low column density across large angular scales, interferometers might miss a significant reservoir of gas around galaxies. Single dishes have full \textit{uv}-coverage, allowing detection of structure at all angular scales. Thus, single dish observations complement interferometric studies of the CGM. 

The unblocked aperture design of the Green Bank Telescope (GBT), decent angular resolution (9.1$'$), its low sidelobes and high surface brightness sensitivity (T$_{sys} \leqslant$ 20 K) make it ideal to search for low column density structures over large scales. Single dish \hi surveys mapping the CGM of nearby galaxies, such as the GBT counterparts to the MeerKAT \hi Observations of Nearby Galaxies; Observing Southern Emitters \citep[MHONGOOSE,][]{Sorgho2019} survey\footnote{\url{https://mhongoose.astron.nl/sample.html}}, and the HALOGAS survey \citep{Pingel2018}, attempted to find diffuse \hi structure around these galaxies, but lacked the sensitivity reached in our work. The upcoming Parkes-IMAGINE (Imaging Galaxies Intergalactic and Nearby Environments) survey\footnote{\url{http://www.imagine-survey.org/imagine-galaxy-sample/}} will map the entire CGM of 28 nearby galaxies in \hin, reaching sensitivities nearly an order of magnitude deeper than other \hi surveys (Sardone et al. 2020 in prep.). 

Previous single dish observations have targeted the CGM of several nearby galaxies. \cite{deBlok2014,Pisano2014} mapped the circumgalactic region of NGC\,2403, NGC\,2997 and NGC\,6946 down to a 5$\sigma$ detection limit of $\sim$10$^{18}$ cm$^{-2}$ over a 20 km s$^{-1}$ channel. The \hi observations of NGC\,6946 revealed a filamentary feature of N(\hin) = 5 $\times$ 10$^{18}$ cm$^{-2}$, apparently connecting the galaxy with its nearest companions. Observations of NGC\,2403 revealed a low column density, extended cloud outside the main \hi disk, either accreting from the IGM or the result of a minor interaction with a neighboring dwarf galaxy. However, the halo of NGC\,2997 did not show any filamentary features, and the \hi mass as measured with the GBT was only 7\% higher than that derived from past interferometric measurements.

The single dish surveys described above integrate for a few minutes per point and cover a large area, usually out to the virial radius, around each galaxy. In this paper, we adopt a complementary approach. We integrate for 0.5--4.5 hours on-source toward individual pointings along the minor axes of two nearby edge-on galaxies: NGC\,891 and NGC\,4565. Our observations are an order of magnitude deeper than earlier single dish observations. We achieve a 1$\sigma$ sensitivity limit of 1.6$\times$10$^{16}$ cm$^{-2}$ calculated over a 20 km s$^{-1}$ line width. These are among the deepest for external galaxies obtained to date in 21-cm \hin. Achieving this sensitivity while also mapping a large area would require a very large time investment. 

Several interferometric and single dish studies have investigated our targets specifically. The WSRT interferometric observations of NGC\,891 revealed a huge gaseous halo extended out to 15 kpc from the disk. This halo contains almost 30\% of the detected \hi \citep{Oosterloo2007}. Using high resolution HALOGAS data, \cite{Zschaechner2012} found the evidence of interaction between NGC\,4565 and its companions, but did not detect any extraplanar diffuse \hi around NGC\,4565. \cite{Pingel2018} compared the GBT observations with the WSRT data at equal spatial and velocity resolution. They did not detect any considerable amount of excess \hi from the GBT data. This indicated that the column density of diffuse \hin, if present, is much lower than their 5$\sigma$ GBT detection limit of 0.9--1.4$\times$10$^{18}$ cm$^{-2}$ over a 20 km s$^{-1}$ channel. 

We outline our observation and the data reduction in section \ref{sec:obsred}. We discuss our analysis and results in section \ref{sec:analysis} and interpret that in section \ref{sec:discussion}. We summarize our conclusions and comment on future directions in section \ref{sec:conclude}.

\section{Observation and data reduction}\label{sec:obsred}
\begin{deluxetable*}{cccccccccc}
\tablenum{1}
\tablecaption{Details of observations and the measurements\label{tab:obs}}
\tablewidth{0pt}
\tablehead{
\colhead{Pointing$^{(a)}$} & \colhead{RA} & \colhead{Dec} & \colhead{t$_{exp}$$^{(b)}$} & \colhead{$\sigma_{T_B}$} & \colhead{d$_{\perp}$ } & \colhead{log$_{10}$N(\hin)$_{GBT}$$^{(c)}$} & \colhead{Mass$^{(d)}$} & \colhead{$v_{avg}$} & \colhead{log$_{10}$N(\hin)$_{WSRT}$$^{(e)}$}
\\
& (J2000) & (J2000) & (hrs) & (mK) & (kpc) & (cm$^{-2}$) & (M$_\odot$) & (km s$^{-1}$) & (cm$^{-2}$)}
\decimalcolnumbers
\startdata
\hline
\multicolumn{10}{c}{\bf NGC\,891} \\
\multicolumn{10}{c}{M(\hin)$_{disk}$=$4.1\times10^9$ M$_\odot$, SFR = $3.92\pm1.75$ M$_\odot$yr$^{-1}$, sSFR = $1.3\times10^{-9}$ yr$^{-1}$, $\Sigma_{SFR}=6.5\times10^{-3}$ M$_\odot$ yr$^{-1}$ kpc$^{-2}$} \\
\hline 
center   & 2$^h$22$^m$33.6$^s$ & 42$^\circ$20$'$58$''$ & 0.10 &7.8 & 0  & 20.701$\pm$0.001 & 3.0$\pm$0.6$\times$10$^9$   & 1.1$\pm$0.1 & 20.672$\pm$0.008\\ 
up\,0.5  & 2$^h$22$^m$08.4$^s$ & 42$^\circ$22$'$44$''$ & 0.33 &3.3 & 13 & 20.370$\pm$0.001 & 1.4$\pm$0.3$\times$10$^9$  & 4.1$\pm$0.1  & 20.393$\pm$0.009\\
down\,0.5& 2$^h$22$^m$58.4$^s$ & 42$^\circ$19$'$00$''$ & 0.33 &2.2 & 13 & 20.306$\pm$0.001 & 1.2$\pm$0.2$\times$10$^9$  & 1.1$\pm$0.1 & 20.344$\pm$0.009\\
up\,1    & 2$^h$21$^m$43.5$^s$ & 42$^\circ$24$'$42$''$ & 0.50 &1.5 & 26 & 19.254$\pm$0.005$^{+0.019}_{-0.045}$ & 1.1$\pm$0.2$\times$10$^8$  & 7.2$\pm$0.5 & 19.250$\pm$0.014\\
down\,1  & 2$^h$23$^m$23.6$^s$ & 42$^\circ$17$'$07$''$ & 0.50 &1.3 & 26 & 19.074$\pm$0.008$^{+0.004}_{-0.062}$ & 7.1$\pm$1.4$\times$10$^7$  & 4.1$\pm$0.7 & 19.195$\pm$0.014\\
up\,1.5  & 2$^h$21$^m$18.7$^s$ & 42$^\circ$26$'$26$''$ & 0.33 &1.8 & 39 & 18.397$\pm$0.047 & 1.5$\pm$0.4$\times$10$^7$  &28.0$\pm$5.0 & 17.952$\pm$0.064\\
         &                     &                       &      &    &    & 18.419$\pm$0.047$\dagger$ & 1.6$\pm$0.4$\times$10$^7$  &  40.2$\pm$6.0 &                 \\
down\,1.5 & 2$^h$23$^m$48.4$^s$ & 42$^\circ$15$'$14$''$ & 2.26 &0.7 & 40 & 17.930$\pm$0.052$^{+0.111}_{-0.139}$ & 4.9$\pm$1.3$\times$10$^6$  &23.4$\pm$6.3 & 17.944$\pm$0.064\\
    &                           &                       &      &    &    & 17.989$\pm$0.049$\dagger$ & 5.9$\pm$1.4$\times$10$^6$ &50.4$\pm$6.9 & \\
down\,2   & 2$^h$24$^m$14.0$^s$ & 42$^\circ$13$'$21$''$ & 3.57 &0.6 & 53 & 18.055$\pm$0.025$^{+0.106}_{-0.068}$ & 6.7$\pm$1.5$\times$10$^6$  &32.7$\pm$4.6  & 17.458$\pm$0.195\\
    &                           &                       &      &    &    & 18.114$\pm$0.024$\dagger$ & 7.8$\pm$1.6$\times$10$^6$  & 63.8$\pm$4.2\\
down\,3   & 2$^h$25$^m$04.0$^s$ & 42$^\circ$09$'$36$''$ & 3.57 &0.4 & 80 & 17.656$\pm$0.062$^{+0.221}_{-0.054}$ & 2.6$\pm$0.8$\times$10$^6$  &-7.6$\mp$7.7 & 17.117$\pm$0.427\\
    &                           &                       &      &    &    & 17.741$\pm$0.055$\dagger$ & 3.3$\pm$0.8$\times$10$^6$  &41.4$\pm$8.0\\
\hline
\multicolumn{10}{c}{Along the minor axis: M(\hin)$_{d,CGM}$ = $5.9\pm0.5^{+0.6}_{-1.0}\times10^7$ M$_\odot$, Extrapolated to whole CGM: M(\hin)$_{d,CGM}$ = $2.2\pm0.1\pm0.4\times 10^8$ M$_\odot$}\\              
            \hline
\multicolumn{10}{c}{\bf NGC\,4565}\\
\multicolumn{10}{c}{M(\hin)$_{disk}$=$7.3\times10^9$ M$_\odot$, SFR = $0.67\pm0.10$ M$_\odot$yr$^{-1}$, sSFR = $1.1\times10^{-11}$ yr$^{-1}$, $\Sigma_{SFR}=6.9\times10^{-4}$ M$_\odot$ yr$^{-1}$ kpc$^{-2}$} \\
\hline 
center   & 12$^h$36$^m$20.9$^s$ & 25$^\circ$59$'$23$''$ & 0.08 &4.4 &  0 & 20.737$\pm$0.001 & 5.5$\pm$0.4$\times$10$^9$   & -0.6$\mp$0.9 & 20.749$\pm$0.004\\  
up 0.5   & 12$^h$36$^m$36.6$^s$ & 26$^\circ$02$'$48$''$ & 0.43 &1.3 & 16 & 20.367$\pm$0.001 & 2.3$\pm$0.2$\times$10$^9$   & 0.7$\pm$0.9 &20.426$\pm$0.004\\  
down 0.5 & 12$^h$36$^m$05.2$^s$ & 25$^\circ$55$'$46$''$ & 0.43 &1.4 & 17 & 20.382$\pm$0.001 & 2.4$\pm$0.2$\times$10$^9$   & -5.4$\mp$0.9 & 20.429$\pm$0.004\\ 
up 1     & 12$^h$36$^m$52.3$^s$ & 26$^\circ$06$'$27$''$ & 0.43 &1.8 & 34 & {18.953$\pm$0.013$^{+0.186}_{-0.104}$} & {9.0$\pm$0.7$\times$10$^7$}   & 14.1$\pm$1.8 & 19.266$\pm$0.008\\  
down 1   & 12$^h$35$^m$49.4$^s$ & 25$^\circ$52$'$13$''$ & 0.43 &1.3 & 34 & 19.128$\pm$0.008$\ddagger$ & 1.4$\pm$0.1$\times$10$^8$   & 1.9$\pm$13.3 & 19.255$\pm$0.006\\  
up 1.5   & 12$^h$37$^m$08.0$^s$ & 26$^\circ$09$'$56$''$ & 3.67 &0.6 & 51 & 18.081$\pm$0.039$^{+0.215}_{-0.006}$ & 1.2$\pm$0.1$\times$10$^7$ & 23.4$\pm$4.3  & 17.982$\pm$0.047\\
         &                      &                       &      &    &    & 18.119$\pm$0.037$\dagger$  & 1.3$\pm$0.1$\times$10$^7$ & 36.8$\pm$4.9 \\
down 1.5 & 12$^h$35$^m$34.0$^s$ & 25$^\circ$48$'$41$''$ & 0.41 &0.2 & 52 & 18.438$\pm$0.046$\ddagger$ & 2.8$\pm$0.3$\times$10$^7$   & -10.3$\mp$59.0 & 17.979$\pm$0.046\\  
up 2     & 12$^h$37$^m$23.7$^s$ & 26$^\circ$13$'$27$''$ & 4.69 &0.5 & 68 & 17.934$\pm$0.046$^{+0.180}_{-0.085}$ & 8.6$\pm$1.1$\times$10$^6$ & 34.2$\pm$6.0 & 17.626$\pm$0.103\\  
up 3     & 12$^h$37$^m$55.0$^s$ & 26$^\circ$20$'$24$''$ & 4.69 &0.4 & 103& 17.582$\pm$0.098$^{+0.177}_{-0.073}$ & 3.8$\pm$0.9$\times$10$^6$& -16.9$\mp$7.2 & 17.291$\pm$0.224 \\ 
         &                      &                       &      &    &    & 17.657$\pm$0.086$\dagger$ & 4.6$\pm$1.0$\times$10$^6$ & 9.1$\pm$6.8  \\
\hline
\multicolumn{10}{c}{Along the minor axis:  M(\hin)$_{CGM}$ = {$3.4\pm0.7\pm0.4\times10^7$ M$_\odot$}, Extrapolated to whole CGM: M(\hin)$_{CGM}$ = {$1.5\pm0.4^{+0.5}_{-0.4}\times10^8$ M$_\odot$}}\\ 
\enddata
\footnotesize{
{$(a)$ The numbers correspond to the offset, e.g., pointing = `up 2' is 2$\times$9.1$'$ away from the center of the target galaxy along the minor axis, where 9.1$'$ is the FWHM size of the GBT beam. ``Up" denotes offset to higher declination, while ``down" denotes offset to lower declination both along the minor axis. The pointings have been illustrated with labels in Figure \ref{fig:NGC891} \& \ref{fig:NGC4565}. The transverse distance of the pointings from the center of the target galaxy are provided in the 6th column.}}

{$(b)$ The exposure times are on-source times. Some of the on-source pointings share a common off-source pointing $\approx 1^\circ$ away from the source. The time spent on each off-source pointing is greater than that in each consecutive on-source pointing. But, the total time spent on the off-source pointing is less than the total exposure time for on-source pointings because of these shared off-source pointings. The total integration time toward the off-source pointing is 7.12 hours for NGC\,891, and 8.60 hours for NGC\,4565.}

{$(c)$ The intensity (in the units of K km s$^{-1}$) is obtained by integrating T$_B$ from -250 to 250 km s$^{-1}$ for NGC\,891 and -300 and 300 km s$^{-1}$ for NGC\,4565. The intensity is multiplied by 1.82$\times 10^{18}$ to convert it to the column density. We assume that the emission is optically thin, and the density of the emitting \hi is well above the critical  density. Here, the error is the combination (in quadrature) of the statistical uncertainty in N(\hin) and the multiplicative systematic uncertainty in flux calibration. {As the emission is 500--600 km s$^{-1}$ broad, we estimate the statistical uncertainty over the same velocity range.} The additive systematic uncertainties obtained from jackknife resampling are included at the pointings wherever applicable. }

{$\dagger$ T$_B$ is integrated from $v=-200$ to $320$ km s$^{-1}$ for NGC\,891 and from $v=-300$ to $360$ km s$^{-1}$ for NGC\,4565, as there are detectable \hi emission in these velocity ranges at these pointings (Figure \ref{fig:NGC891} and \ref{fig:NGC4565})}

{$\ddagger$ At 0--250 km s$^{-1}$, these pointings are dominated by the emission from NGC\,4562, a nearby companion southwest of NGC\,4565 (Figure \ref{fig:NGC4565}, yellow area). To calculate the intensity at these pointings, we assume that the emission from NGC\,4565 is approximately flat over the whole velocity range, including that obscured by NGC\,4562. We integrate T$_B$ from -300 to 0 km s$^{-1}$ and 250 to 300 km s$^{-1}$, then linearly scale it to the range of -300 to 300 km s$^{-1}$. In this calculation, we have assumed that the spectra are not contaminated by NGC\,4562 beyond 0--250 km s$^{-1}$. Extending this velocity range to -25 to 275 km s$^{-1}$ and -50 to 300 km s$^{-1}$ did not change the intensity significantly, thus validating our assumption.}

{$(d)$ The errors include the multiplicative systematic uncertainty in flux calibration as well as the statistical uncertainty.}

{$(e)$ The errors include the systematic uncertainty in the masking threshold as well as the statistical uncertainty.}
\end{deluxetable*}

\noindent We observed the 21-cm (1.42 GHz) line  emission from the halo of NGC\,891 and NGC\,4565 on 9--12th October, 2015 and 12th June--4th July, 2016 as part of the GBT project 15B-257. NGC\,891 is at a distance of 9.2$\pm$0.9 Mpc \citep{Mould2008}, and NGC\,4565 is at a distance of 11.9$\pm$0.3 Mpc \citep{Radburn-Smith2011}. At these distances, the $9.1'$ GBT beam translates to $24.4-31.5$ kpc. This is large enough that we expect to sample a representative volume of CGM.
This allows us to use multiple pointings to map out the structure of the CGM. To avoid the extended \hi disk contamination along and around the major axis, we chose the pointings along the minor axis, which can be done only around sufficiently inclined galaxies. Our galaxies are highly inclined; with $i \geqslant 88.6^\circ$ for NGC\,891 and $i \geqslant 87.5^\circ$ for NGC\,4565 \citep{Rupen1991}. 

Our initial observation consisted of five pointings (Figure \ref{fig:NGC891} and \ref{fig:NGC4565}, solid circles) separated from one another by the FWHM beam size. We supplemented these full beam-spaced pointings with additional pointings at intermediate positions, i.e., separated by half the FWHM beam size from the other pointings (Figures \ref{fig:NGC891} and \ref{fig:NGC4565}, dashed circles).  The details of the observations, including the sky direction of the pointings and the integration time at each pointing are provided in Table \ref{tab:obs}. 

We observed by position switching, using the L-band receiver with the Versatile GBT Astronomical Spectrometer (VEGAS; bandwidth = 23.5 MHz) as the backend. We used the quasars 3C\,48 and 3C\,286 as the primary flux calibrators for NGC\,891 and NGC\,4565, respectively. The disk of each galaxy was observed before the observations of the CGM to verify the setup.

We adopted an aperture efficiency at 1.42 GHz of 0.6575 and assumed an atmospheric opacity at zenith to be 0.01 to calculate the effective temperature of the noise diode, T$_{cal}$\footnote{\url{http://library.nrao.edu/public/memos/gbt/GBT_289.pdf}}. Combining these assumptions with the observations of the flux calibrators, we found that T$_{cal}$ varies between 1.38 K and 1.58 K for the XX polarization and between 1.54 K and 1.71 K for the YY polarization with a 1$\sigma$ scatter of 0.09 K and 0.07 K on the mean values of 1.49 K and 1.60 K, respectively. When the polarizations are averaged to derive the Stokes\,I component, this translates to $\pm 4.4\%$ systematic uncertainty due to uncertain flux calibration, and $T_{\rm sys}$ variation. The typical system temperature, T$_{sys}$, during our observations fell between 17.4 K and 22.1 K.

We developed a routine in GBTIDL to reduce our position-switched data\footnote{\url{http://gbtidl.nrao.edu}}. Using this routine we extract the calibrated, off-subtracted spectrum at each pointing in each session for both polarizations in units of brightness temperature, T$_B$. We shift the velocity axis of each spectrum so that $v = 0$ km s$^{-1}$ corresponds to the systemic velocity of the galaxy, $v_{sys}$. This is 528 km s$^{-1}$ for NGC\,891 and 1230 km s$^{-1}$ for NGC\,4565 \citep{Rupen1991}. 

We fit and subtract baselines to the spectrum near the emission from each galaxy. For NGC\,891, we use the velocity range of $-400\leqslant v - v_{sys} < -250$ km s$^{-1}$ and $250< v - v_{sys} \leqslant 700$ km s$^{-1}$ to estimate the baseline. For NGC\,4565, we use the velocity range of $-750\leqslant v - v_{sys} < -300$ km s$^{-1}$ and $300< v - v_{sys} \leqslant 700$ km s$^{-1}$ to estimate the baseline. Thus, the velocity ranges used for baseline estimation (600 km s$^{-1}$ and 850 km s$^{-1}$) are broader than the velocity range where we expect emission (500 km s$^{-1}$ and 600 km s$^{-1}$). The outer limits of the velocity ranges used for the fit are set such that beyond this range the baseline is too wavy to obtain a decent fit with a low-order polynomial. We also choose the range to avoid emission from the Milky Way. The inner limits of the velocity ranges are set by the velocity range of the disk of the galaxies. We assume that the emission at all pointings falls within the same velocity range as the disk. For the pointings close to the halo ($\lesssim 40$ kpc), we fit the baselines using first order polynomials. For the pointings farther out, the baselines vary from session to session; we use first through third order polynomials to fit them, with the decision made after by eye inspection. We accumulate the raw baseline-subtracted spectra in individual session at the native velocity resolution of 1.25 km s$^{-1}$.  

For each pointing, we combine the spectra for each polarization separately from all relevant sessions weighting by {T$_{sys}$ and} the exposure time, t$_{exp}$\footnote{Because $\sigma_{T_B} \propto \frac{ {T_{sys}}}{\sqrt{t_{exp}}}$, weighing T$_B$ with t$_{exp}$ {and T$_{sys}$} implies that each session is treated equally, {At any pointing, the scatter of $T_{sys}$ is only $2-3\% (1-2\%)$ for NGC\,891 (NGC\,4565).} }. Thus we obtain the mean T$_{XX}$ and T$_{YY}$. We compare the XX and YY spectra at each pointing to look for systematic effects, and find that they are consistent with each other. Finally, we average the two polarizations to increase the S/N by a factor of $\approx \sqrt{2}$. After averaging, we fit all spectra using 0th--3rd order polynomials to remove any residual baseline. Finally, we boxcar smooth the baseline-subtracted spectra to a final velocity resolution of 20 km s$^{-1}$. 

As a check, we compare the RMS noise estimates from the signal-free part of the smoothed spectrum at each pointing ($\sigma_{T_B}$ in Table \ref{tab:obs}) to the noise expected based on the radiometer equation. We find that the RMS noises are higher than the expected theoretical noises by a factor of $\approx 1.0-1.6$ (except at the pointings `center' and `up0.5' around NGC\,891, where the observed noise is 2.4-2.5 times the expected noise. However, as our focus is not on these points, this discrepancy does not become an issue in our analysis).

As we combine the spectra from different sessions, the difference in the baseline subtraction from one session to another may add a systematic uncertainty in addition to the statistical uncertainties. At the farthest pointings (up/down 1.5, 2, 3), we perform a jackknife sampling to determine any bias. These measurements for individual sessions appear as sky-blue points in Figure \ref{fig:column}, and are consistent with the averaged value within $\pm 1.9\sigma$. This confirms that the detections are not biased by the values of any single session. We consider the scatter among the individual sessions to capture an alternative source of uncertainty, and quote it as an additive systematic uncertainty. However, it should be noted that the scatter between sessions does include a component of statistical uncertainty and the multiplicative systematic uncertainty due to T$_{cal}$ variation. Therefore, we may modestly overestimate our uncertainty.

\section{Analysis and results}\label{sec:analysis}
\begin{figure*}
        \centering
        \includegraphics[trim = 5 0 45 27, clip,scale=0.475]{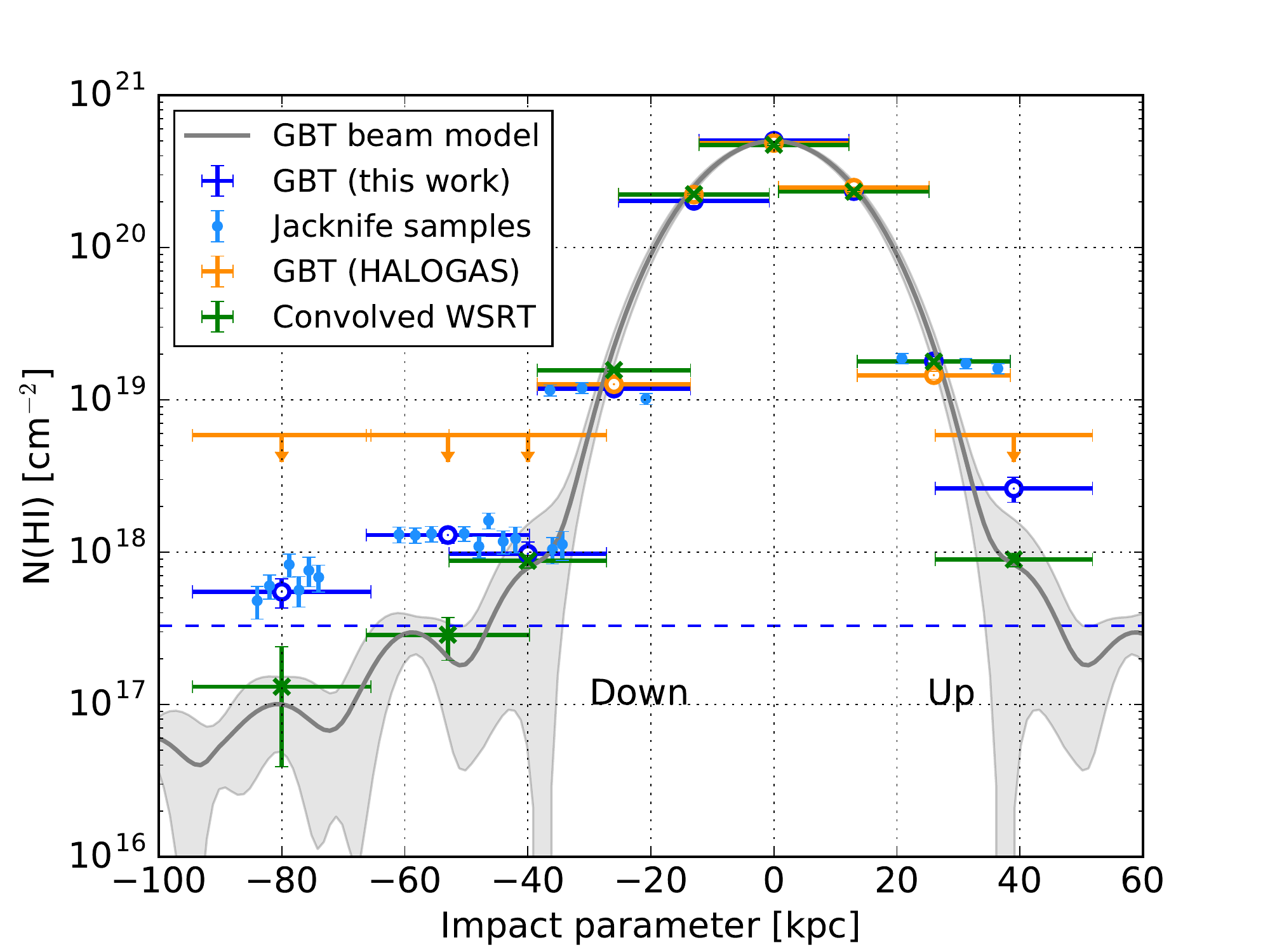}
        \includegraphics[trim = 5 0 45 27, clip,scale=0.475]{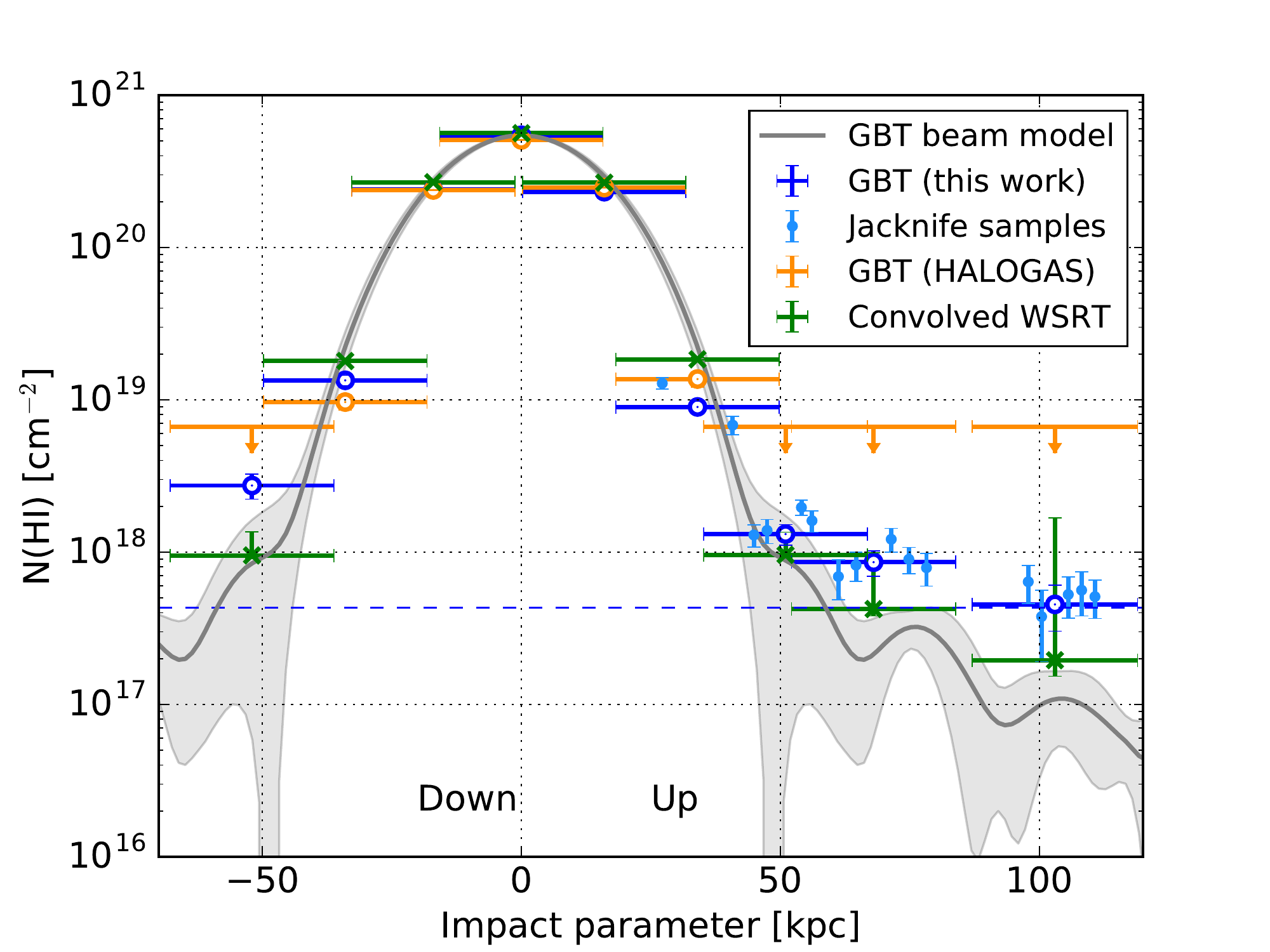}
    \caption{The radial N(\hin) profile of NGC\,891 (left) and NGC\,4565 (right). The error bars include systematic uncertainties due to flux calibration (our GBT data) or the difference in masking threshold (S/N$>$3.5--5--6.5 in the WSRT data), and correspond to 90\% confidence intervals. The upper limits are 5$\sigma$ limit from HALOGAS calculated over a 500--600 kms$^{-1}$ linewidth. The dashed horizontal line is the 5$\sigma$ sensitivity of our observation calculated over a 500--600 kms$^{-1}$ velocity range. The shaded region around the beam model is 1$\sigma$ uncertainty in the beam response due to averaging the beam map to one dimension. The positive offset between our GBT analysis (blue points) and the convolved WSRT data (green points) shows the tentative detection of extended diffuse \hin.}
    \label{fig:column}
\end{figure*}
\noindent In the final spectra, we see clear detections of \hi emission in most of the pointings (Figure \ref{fig:NGC891} \& \ref{fig:NGC4565}). The velocity width of these emission profiles resembles that of the disk of the galaxies, as seen from the central pointings (Figure \ref{fig:NGC891} \& \ref{fig:NGC4565}, gray area). We integrate each spectrum from -250 to 250 km s$^{-1}$ for NGC\,891 and -300 to 300 km s$^{-1}$ for NGC\,4565 to obtain the intensity in units of K km s$^{-1}$. {Because the emission is 500--600 km s$^{-1}$ broad, we estimate the statistical uncertainty in the intensity over the same velocity range.} Also, we calculate the mean velocity over the same velocity range using $v_{avg} = \frac{\int T_Bv dv}{\int T_B dv}$. 

We convert the integrated intensity at each pointing to column density, N(\hin), assuming that the emission is optically thin and the density of the colliders exciting the \hi is well above the critical density\footnote{At a temperature of 10$^4$ K where the CGM is predominantly ionized, the critical density of the 21-cm line for collisions with electrons is 7$\times$10$^{-6}$ cm$^{-3}$ \citep{Draine2011}. The lowest N(\hin) in our GBT measurement translates to a total hydrogen column of N(H) = 3.8$\times$10$^{19}$ cm$^{-2}$ for a neutral hydrogen fraction of 0.01 \citep{Popping2009,Bland-Hawthorn2017}. Assuming a path length of the order of the virial radius $\approx$250 kpc, the average gas density is 3$\times$10$^{-5}$ cm$^{-3}$, a factor of 4 higher than the critical density.}. We do not apply any correction for the inclination angle in computing the column density, because the geometry of the CGM emission is unknown. We calculate the \hi mass per beam from the respective integrated intensities\footnote{using the equation from \url{https://science.nrao.edu/facilities/gbt/proposing/GBTpg.pdf}} adopting a gain of 1.86 K Jy$^{-1}$. We quote the N(\hin), the associated \hi masses, and v$_{avg}$ in Table \ref{tab:obs}.  

We show the derived column densities as a function of impact parameter in Figure \ref{fig:column} (blue points). Beyond 40--50 kpc, we detect \hi at 5--18$\sigma$ significance, where $\sigma$ is the statistical uncertainty.  After including the uncertainty in flux calibration (see \S\ref{sec:obsred}), the significance reduces to 5--13$\sigma$. The spread among the jackknife samples provides an additional systematic uncertainty at the farthest pointings; we list them in Table \ref{tab:obs}. The column density monotonically decreases with impact parameter and is approximately symmetric between the ``up'' and ``down'' directions. The N(\hin) profiles of NGC\,891 and NGC\,4565 appear very similar to one another at similar angular scales.  

Our derived N(\hin) values agree well with previous, shallower GBT observations obtained in the HALOGAS survey \citep[Figure \ref{fig:column},orange points,][]{Pingel2018}. Our detections at large impact parameter ($\gtrsim 40$kpc) lie below the $5\sigma$ sensitivity limit of their GBT cubes, calculated over 500--600 km s$^{-1}$ velocity range. 

\subsection{Comparing to interferometric data}\label{sec:implication}
{
\noindent We compare our GBT measurements with sensitive interferometric maps from the HALOGAS survey using WSRT. These interferometric data are very sensitive to any compact emission, but they lack our single-dish sensitivity to extended structure. Our GBT observation at any pointing includes both the high column \hi clumps and the diffuse large-scale structures in the CGM, if any, as well as the stray light from the disk of the galaxy.}

{To carry out a rigorous comparison, we convolve the masked interferometric data cubes with the GBT beam. This moves the interferometric data to the angular resolution of the GBT. From the convolved interferometric maps, we can estimate the contribution of compact, relatively high-column structures to our observed GBT spectra. The convolved interferometer map will reflect any emission detected in the interferometer cube, i.e., any bright, compact CGM structure, extraplanar emission, and pickup of the disk of the galaxy from the extended wings of the GBT beam. The excess in the GBT spectra compared to the spectra from convolved WSRT cubes will correspond to the extended structures missed by the interferometer.}

{For this exercise, we perform the following steps: 
\begin{enumerate}
    \item \textbf{Noise estimation:} In the interferometric maps, we do not see any emission in the region beyond $r=20'$ from the center of the target galaxy. So, we extract the pixel values beyond this radius in all velocity channels and use these to determine the noise. We estimate a single global noise value, $\sigma_g$, by fitting the negative half of the histogram of those pixel values with a Gaussian.
    \item \textbf{Primary-beam correction:} We use the MIRIAD task \textit{pbcor} to correct for the primary beam-response of WSRT in the interferometric data cubes\footnote{{The primary-beam correction factor of WSRT increases with the distance from the disk of the galaxy. This leads to strongly spatially variable noise in the primary-beam corrected interferometric data cubes. Therefore we estimate the noise and create the mask before correcting for the primary-beam response.}}. 
    \item \textbf{Masking:} We are interested in emission securely detected by the interferometer. To identify such emission, we set a masking threshold and remove all emission in pixels with S/N below that. The  interferometric cubes masked with S/N$\geqslant5$ are used to calculate the best estimate of the spectra. We vary the masking threshold from S/N$\geqslant3.5$ to S/N$\geqslant6.5$ to estimate the systematic uncertainty associated with this step.
    \item \textbf{Circularization of GBT beam-model:} The GBT beam is not a sharp tophat with width 9.1$'$. It is not azimuthally symmetric either \citep[e.g,][]{Pingel2018}. As the sky position of a given pointing changes during the course of an observation, the orientation of the GBT dish also changes. To account for that, we take the azimuthal average of the GBT beam response. The circularized GBT beam model along with its 1$\sigma$ uncertainty from \cite{Pingel2018} is shown in Figure \ref{fig:column}. Even separated by more than a FWHM, the GBT beam picks up emission from nearby bright sources, including the disk of the target galaxy. 
    \item \textbf{Convolution:} For each pointing, we convolve the primary-beam corrected, masked WSRT cube with the circularized GBT beam-model\footnote{Because of the finite beam size of WSRT, the angular resolution of the convolved WSRT cubes will be larger than that of GBT. However, the beam size of WSRT \citep[$\sim 0.5'-0.7'$,][]{Heald2011} is much smaller than the beam size of GBT ($\sim 9.1'$). Therefore, the difference of angular resolution between the convolved WSRT cubes and our GBT observations is $<1\%$, which is negligible.}. We set the center of the circularized GBT beam-response at the sky location of each GBT pointing (Table \ref{tab:obs}), and treat any missing or masked data in the WSRT cube as having zero intensity. The resulting spectrum represents what we would expect in the GBT spectrum if GBT observed only the securely detected emission in the WSRT cube. This includes stray light from the disk of the galaxy, extraplanar emission and any circumgalactic cloud with high enough column density to be detected by the interferometer (Figure \ref{fig:column}, green points). 
    \item \textbf{Uncertainties:} The statistical uncertainty in the convolved WSRT cube, $\sigma_{con}$ is calculated from the global noise of the unmasked unconvolved WSRT cube, $\sigma_g$ (see step\,1) using the response distribution of the GBT beam across the sky: $\sigma_{con} = \sigma_g \times  \frac{\sqrt{\Sigma w^2}}{\Sigma w}$, where \textit{w} is the GBT beam response. The systematic uncertainty, as mentioned in step\,3, comes by convolving the primary-beam corrected WSRT cubes masked with thresholds from S/N$\geqslant3.5$ to S/N$\geqslant6.5$. 
\end{enumerate}
}

\begin{figure}[h]
    \centering
    \includegraphics[trim= 5 0 40 30,clip,scale=0.45]{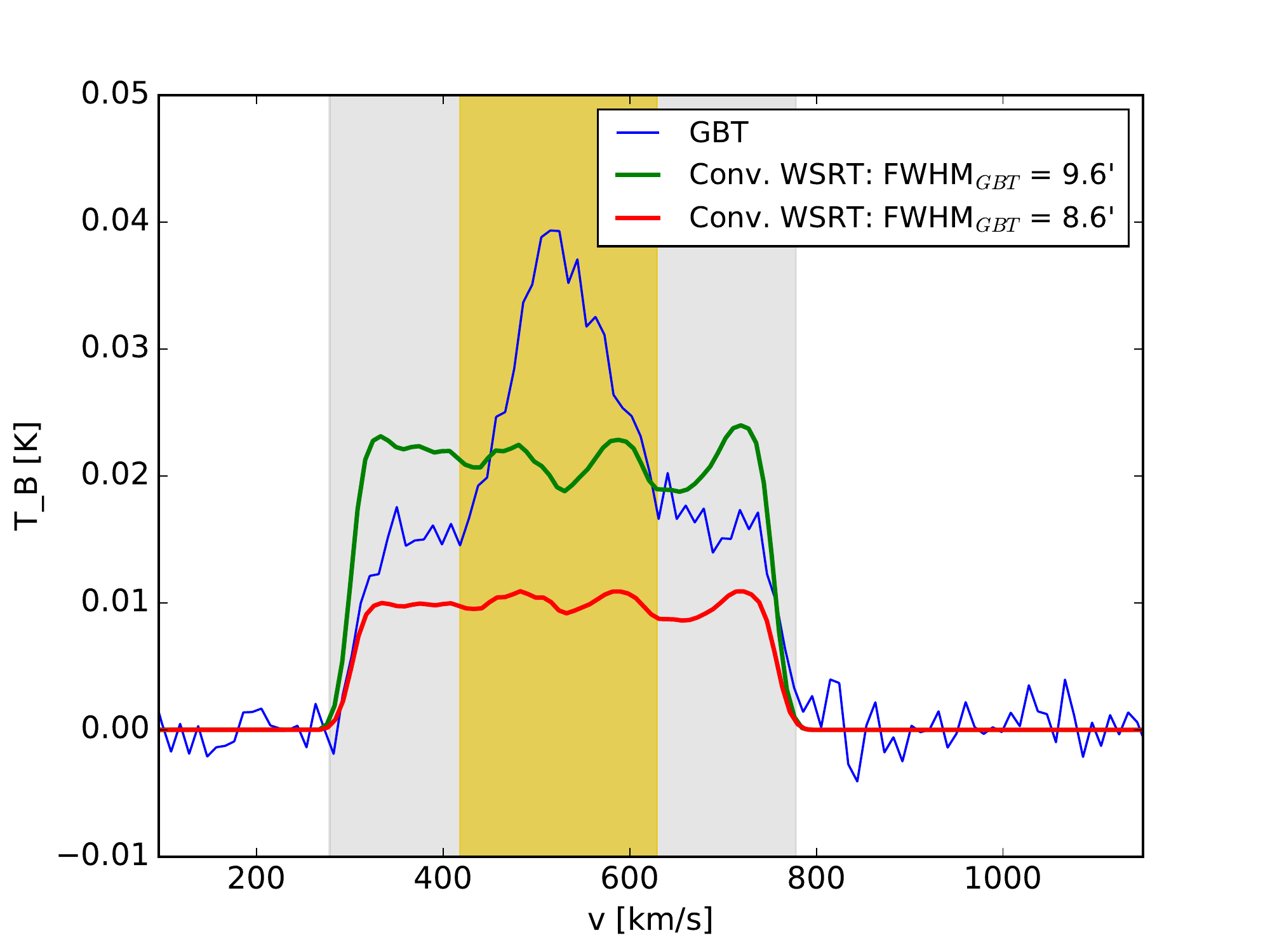}
    \caption{The GBT spectrum of the pointing ``up1" around NGC\,891 (see Figure \ref{fig:NGC891}), and the WSRT spectra convolved with two beam models with FWHM $8.6'$ and $9.6'$. Because the $9.6'$ beam reproduces the GBT spectra at the center of the galaxies better than the $8.6'$ beam, we use the $9.6'$ beam in our analysis. The triangular emission in the region shaded in yellow is a suggestive evidence of diffuse, extended \hin.    }
    \label{fig:beamwidth}
\end{figure}

We compare the shape and the integrated intensity of our GBT spectrum to the convolved WSRT spectrum at each pointing. We find three sets of results: \\
\begin{enumerate}
    \item {\textbf{Inner halo ($r_\perp\leqslant20-30$ kpc)}: Here the \hi emission is dominated by the disk in both the GBT and WSRT spectrum. The WSRT measurements are comparable with those from the GBT.}
    \item {\textbf{Large impact parameters (pointings up/down 1.5, 2 and 3; $r_\perp\geqslant40-50$ kpc)}: There is a significant excess in our GBT measurements compared to those from the WSRT (Figure \ref{fig:column}). The WSRT columns account for $42\pm20$\%($50\pm11$\%) of the GBT columns of NGC\,891 (NGC\,4565).\\ Note that the column densities at these pointings are below the sensitivity limit of WSRT ($\approx 10^{19} cm^{-2}$). Therefore, the masked WSRT cube does not have any high significance emission within $\approx9.1'$ (the FWHM of the GBT beam) of these pointings. That means the masked and convolved WSRT spectra at these pointings represent the stray light from the \hi disk and the immediate extraplanar region. Therefore, we infer that the offset between our GBT and WSRT columns at these pointings represents the diffuse \hi emission from the CGM. \\
    This diffuse emission could consist of a mixture of small \hi clumps/clouds, and the diffuse, extended \hin, as is evident from the spectral shape of GBT at these pointings (see figure \ref{fig:NGC891} and \ref{fig:NGC4565}). The large line width of the spectrum implies that whatever the form, this emission must be widespread along the line of sight. Deeper interferometric data are necessary to distinguish between the small scale and extended \hi structures at these pointings.}
    \item {\textbf{Intermediate impact parameters (pointings up/down 1; $20-30\leqslant r_\perp\leqslant40-50$ kpc)}: Because the column density of \hi sharply drops off near pointings ``up1/down1" in both galaxies, the convolved WSRT spectra at these pointings are extremely sensitive to the adopted GBT beam model. Here the assumption of a circular beam might not be adequate. Despite this uncertainty, the shape of the spectrum gives some clue as to the origin of the observed \hin. The contribution of the \hi disk at any of our extended pointings shows a spectrum that resembles the disk spectrum. Mainly the amplitude, not the shape, depends on the details of the beam response. According to this criterion, at pointing ``up1" around both galaxies and at pointing ``down1" around NGC\,891, we find that the emission cannot be solely explained by the disk contamination.} \\
    In Figure \ref{fig:beamwidth}, we plot the GBT spectrum at pointing ``up1" around NGC\,891, and compare it with the WSRT spectra convolved with two different GBT beam models, one with FWHM $8.6'$ and one with FWHM $9.6'$. Both WSRT spectra show 500 km/s broad emission (shaded in gray) from the \hi disk. The GBT spectrum shows a similar component, with an amplitude between that from the $8.6'$ and $9.6'$ GBT beam models. This indicates that the true GBT beam width must be somewhere in between $8.6'$ and $9.6'$. The shape and the column densities of the GBT and the convolved WSRT spectra at the center of the galaxies are similar for the $9.6'$ beam, so the discrepancy is unlikely due to calibration offset between the single dish and the interferometer. Also, as the beam response is not azimuthally symmetric, circularizing the beammaps might affect the spectral shape and amplitude at these sensitive pointings. \\
    Our GBT spectrum shows a 200 km/s wide emission (shaded in yellow) in addition to the 500 km/s wide emission produced by the convolved WSRT.
Because the interferometric data have limited sensitivity to any smooth, extended \hi component, we interpret this excess in the GBT compared to the WSRT as the emission from the diffuse, extended \hi structures. 
\end{enumerate}

\subsection{Comparing to absorption studies}\label{sec:absorption}

\noindent We compare our emission-based measurements with absorption measurements from the literature. From the Lyman series analysis of 44 COS-Halos systems, \cite{Tumlinson2013} set an upper limit of 10$^{19}$ cm$^{-2}$ and a lower limit of 10$^{16}$ cm$^{-2}$ on the \hi column densities. The large range reflected the difficulty of extracting column densities from Ly-$\alpha$ absorption in this regime, when the damping wings are not yet strong but the lines are optically thick. Our column densities at r$\gtrsim$40 kpc are in the range of $5\times10^{17}-3\times10^{18}$ cm$^{-2}$, in good agreement with \cite{Tumlinson2013}. 

\begin{figure}[h]
    \centering
    \includegraphics[trim= 5 0 40 30,clip,scale=0.45]{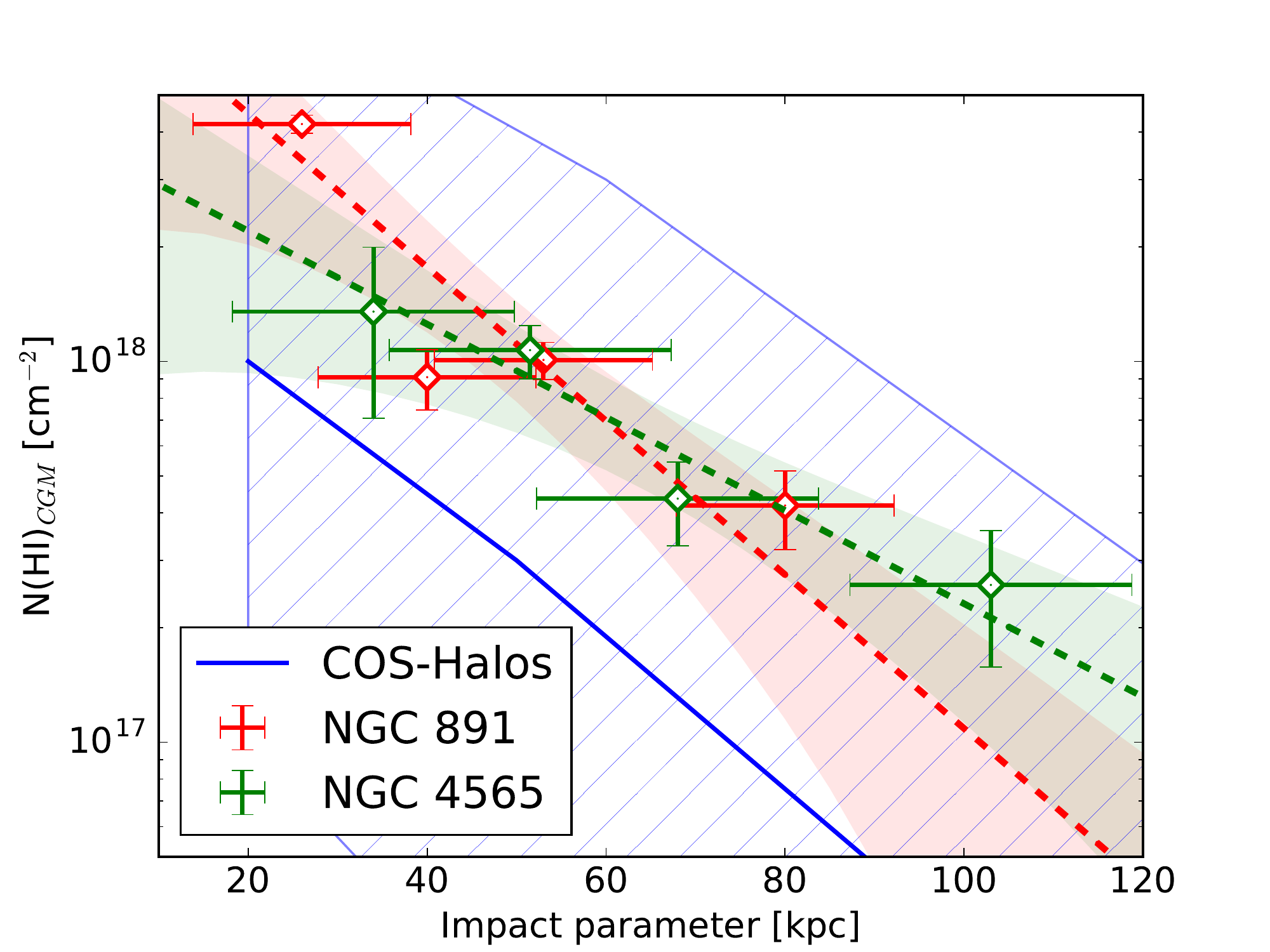}
    \caption{The profile of diffuse, extended \hi column density derived by subtracting the convolved WSRT integrated intensity from the GBT integrated intensity, and then averaging the two sides of the disk. The hatched area is the range of N(\hin) measured in the COS-Halos sample of star-forming galaxies \citep{Prochaska2017}. COS-Halos does not include absorption measurements with impact parameters $<$20\,kpc. The areas shaded in red and green correspond to our best-fit profiles with scale lengths of 22$\pm$7\,kpc (NGC\,891) and {36$\pm$8\,kpc} (NGC\,4565).}
    \label{fig:diffuse}
\end{figure}

Based on more detailed modeling, \cite{Prochaska2017} estimated the N(\hin) of 14 COS-Halo Lyman limit systems. The median and range of their estimated typical N(\hin) profile are shown in Figure \ref{fig:diffuse} (blue curve and hatched area). A direct comparison between our emission based results and the absorption work by \cite{Prochaska2017} is complicated by the different targets and distinct nature of the observations. The N(\hin) profile of galaxies is not necessarily independent of their galactic properties, and may not be azimuthally symmetric. The COS-Halo sample does not include our targets, and spans two orders of magnitude in stellar mass and specific star formation rate (sSFR). \cite{Tumlinson2013} and \cite{Prochaska2017} do not take the relative orientation of galaxy-quasar pairs into account in their analysis, so we cannot restrict our comparison to the minor axes of COS-Halos galaxies.
All of this means that we can only compare average trends in the COS-Halo sample with our targets.  Also, the absorption lines sample a ``pencil beam'' through the CGM, while the emission measurements convolve a large area, including potentially stray light pickup from the disk (see above). A direct comparison of absorption with our GBT measurement must take into account this different spatial sampling, which is not trivial. 

As a zeroth order comparison, we subtract the convolved WSRT values from our GBT values at each pointing. When doing so, we propagate the statistical and systematic errors of each. If this excess does come from the CGM including the diffuse, extended \hi and small clouds/clumps, we might expect this to also be the column density sampled by absorption measurements.
Unfortunately, imperfect knowledge of the beam model and azimuthal structure leave uncertainties in our subtraction of the WSRT. For the pointings ``up1/down1", the calculation is more sensitive. Therefore, we make a rough estimate, considering only the 200 km/s wide line sitting above the flat component as discussed in the previous section. By approximating the shape of the line as a triangle, we estimate the N(\hin)$_{diff} \sim 5.0 (3.4) \times 10^{18}$ cm$^{-2}$ for pointing ``up1 (down1)" around NGC\,891 and  N(\hin)$_{diff} \sim$ {1.4$\times 10^{18}$} cm$^{-2}$ for pointing ``up1"  around NGC\,4565. 

Figure \ref{fig:diffuse} shows that the column density of the \hi that we detect in emission is generally larger than the median absorption seen by COS-Halos at similar impact parameter. However, our measurements do lie well within the maxima of absorption. Therefore, our GBT measurements are roughly consistent with those from Lyman limit \hi absorption. By fitting the profile with an exponential: N(\hin) = N(\hin)$_o$exp(-r/r$_s$), we obtain the scale heights r$_s$ to be 22$\pm$7\,kpc (NGC\,891) and {36$\pm$8\,kpc} (NGC\,4565). This shows that the profile of diffuse \hi in the CGM is steeper in NGC\,891 compared to that in NGC\,4565. Also, we fit the galaxies together, and get an average r$_s$= 28$\pm$6\,kpc. 
\section{Discussion}\label{sec:discussion}
\noindent Using deep GBT observations targeting the minor axes of NGC\,891 and NGC\,4565, we have obtained 6--7$\sigma$ detections of the \hi 21-cm line out to impact parameters as large as 90--120 kpc. By comparing our single dish measurements with interferometric studies, we find that the GBT detects \hi in the CGM of NGC\,891 and NGC\,4565 out to 90--120\,kpc from the galaxies. Also, there is suggestive evidence of diffuse, extended \hi closer to the galaxy, at impact parameter $\approx 30$\,kpc, which may be missed by interferometric observations. Below, we calculate the implied neutral gas content of the CGM and discuss possible origin(s) of the detected \hin.

\subsection{Origin}
\noindent The \hi that we detect in the CGM could originate from the \textbf{IGM}, tidal \textbf{interactions} with other galaxies, \textbf{dwarf satellite/companion galaxies},  cold \textbf{outflows}, or \textbf{condensation} of cold gas from the hot CGM. Below, we consider each possibility. 

\textbf{Emission from companion galaxies:} The \hi mass within the GBT beam is often a few times $10^6-10^7$ M$_\odot$ (Table \ref{tab:obs}). This is comparable to the \hi mass of a dwarf galaxy. However, the linewidths that we observe at those pointings (Figure \ref{fig:NGC891} and \ref{fig:NGC4565}, gray areas) are comparable with the velocity range of the target galaxies, which is also close to the velocity range expected for random motions in the halos of the galaxies. The spectrum of a dwarf companion (e.g., Figure \ref{fig:NGC4565} the yellow area shows the spectrum of a dwarf companion to NGC\,4565) would be noticeably narrower. We would also expect the deep interferometric observations to detect any reasonably compact emission like that expected for a dwarf galaxy. This disfavors the possibility of confusion with a satellite galaxy.

\textbf{Tidal interactions:} NGC\,891 has a filament toward its companion UGC\,1807 in the northwest \citep{Oosterloo2007}. NGC\,4565 has two very nearby companions: NGC\,4562 to the southwest, IC\,3571 directly to the north \citep{Zschaechner2012}. The filament and the companion galaxies are captured in our pointings (Figure  \ref{fig:NGC891}). If the diffuse \hi traces the interaction between the galaxies, the column density might be biased in the direction of the companions. However, the column density profiles are quite symmetric in two sides of the disks along the minor axes, lacking any noticeable excess toward the companions (Figure \ref{fig:column}). {Also, tidal features are often compact enough to be detectable in interferometer maps (e.g., the low column density (N(\hin)$\approx 10^{17} cm^{-2}$) \hi filament between M31 and M33 \citep{Braun2004,Lockman2012} was found to be clumpy and made up of small $\approx$ kpc-scale \hi clouds in high resolution observations by \cite{Wolfe2013}).} Given the morphology of the detected emission, symmetry of the profiles and the low likelihood of lining tidal features exactly along the minor axis in both sides of the disk, we consider it unlikely that these observations are driven by tidal interactions. 


\textbf{Inflow/outflow/ galactic fountain:}
We chose the minor axis to avoid contamination from the extended \hi disk. However, outflows and galactic fountains tend to be aligned with the minor axis, which could lead to preferential enhancement of column density along the minor axis. In principle, this can be checked by looking for velocity gradients or a correlation with the launching mechanism (SFR or AGN). 

There is no appreciable difference between the column density profiles of the galaxies (Table \ref{tab:obs}, Figure \ref{fig:column}). Given that NGC\,891 appears to have a much higher sSFR (specific star formation rate) and $\Sigma_{SFR}$ (surface density of SFR) than NGC\,4565 \citep[Table \ref{tab:obs},][]{Pingel2018}, we might expect NGC\,891 to launch more powerful outflow and/or show a more prominent signature of condensation. However, NGC\,4565 hosts a LINER type AGN \citep{Veroncetty2000}, which is a potential contributor to the outflow. Because the combined effect from star formation and AGN is quantitatively uncertain, the relation between the detected \hi and outflow/ precipitation based on feedback argument remains inconclusive. 

The metal content of the gas may help constrain its origin. From the UV absorption analysis of the warm halo along two sightlines 5 and 108 kpc above the disk near the minor axis of NGC\,891, \cite{Qu2019} infer metallicities of 0.5 and 0.1Z$_\odot$, respectively. We cannot eliminate the possibility of a cold metal-enriched outflow from this. However, the difference in the metallicity at 5 and 108 kpc shows that the metal-enriched gas, even if ejected from the disk, has not reached a large distance. 

Also, we try to infer the origin of the gas from kinematic information. In Figure \ref{fig:NGC891} and \ref{fig:NGC4565}, we have plotted the brightness temperature-weighted average velocity with respect to the systemic velocity of the galaxy along our line of sight. Because the galaxies are edge-on, the observed velocity would represent a small component of the true velocity for any motion perpendicular to the disk. If the dominant component of the emission is a quasi-static medium spread evenly across the halo, the average velocity should be consistent with the systemic velocity of the galaxy. If the observed emission is from the far side of the galaxy, positive (or negative) velocity would imply outflow (or accretion/precipitation). Similarly, positive (or negative) velocity of the emission from the near side of the galaxy would imply accretion/precipitation (or outflow). If a bipolar outflow or inflow is present, we should expect to see a velocity gradient with positive values in one side, and negative values in the other side of the disk.

We observe such gradient in the halo of NGC\,4565 (Table \ref{tab:obs}). However, as we do not know the geometry and the position of the observed gas, we cannot determine whether it is an outflow or an inflow.

In the halo of NGC\,891, the velocity is positive at all pointings and usually increases in magnitude with impact parameter (Table \ref{tab:obs}). Outflow or inflow alone cannot explain such trend. Therefore, it could be produced by a combination of outflow, inflow, and galactic fountain activity. This is consistent with the observation and modeling of NGC\,891 \citep{Oosterloo2007,Fraternali2008}, where a galactic fountain is inferred from the increasing lag in rotation with height. It is also possible that the mean velocity of the disk of NGC\,891 is slightly lopsided, so that it is not perfectly representative of the halo mean velocity. However, such warped \hi profiles are actually visible in both of our targets, with the stronger in NGC\,4565 \citep{Rupen1991,Oosterloo2007,Zschaechner2012}. Though not conclusive, it would be unexpected if the less prominent warp in NGC\,891 affected the kinematic alignment significantly.

\textbf{Condensation:} Note that the absolute values of the average velocities are within 10\% of the velocity width, i.e., $|v_{avg}| < 0.1\times v_{width}$. That is indicative of a quasi-static neutral medium permeating the halo in addition to any inflow, outflow and galactic fountain. In the ``hot mode" accretion common for galaxies as massive as our targets, the infalling gas from the IGM becomes shock-heated and virialized and forms a quasi-spherical hot halo \citep{Rees1977,Keres2005}. The diffuse \hi we detect could be a neutral phase mixed with this predominantly hot medium, which would eventually cool and accrete onto the galaxy.  

Previous studies have explored extraplanar emission from both of our targets at other wavelengths. The studies find bright radio continuum, dust filaments, H$\alpha$ recombination and X-ray emission in the extraplanar region of NGC\,891 but no corresponding features in NGC\,4565 \citep{Rand1992,Howk1997,Rand1998,Howk1999,Strickland2004a,Tullmann2006,Hodges-Kluck2018,Schmidt2019}. These differences have been attributed to the stark difference in the star formation activity of these galaxies. This also suggests that the \hi halos of NGC\,4565 and NGC\,891 may have different origins. 

All of these studies focused within 10\,kpc of the disk. It is not clear if the effect of disk-halo interaction remains similar out to the large impact parameters as probed by our observations. The \hi profile is flatter in NGC\,4565 (\S\ref{sec:absorption}), with smaller \hi column within 40\,kpc (Figure \ref{fig:diffuse}), as compared to NGC\,891. This implies that despite having similar \hi masses within the volumes probed (Table \ref{tab:obs}), in the inner CGM NGC\,4565 has less \hi mass than NGC\,891 by a factor of a few. It is worth noting that the SFR of NGC\,4565 is smaller than the SFR of NGC\,891 by about the same factor as the \hi mass difference \citep{Pingel2018}. If our detected \hi is from the condensing hot CGM accreting to the galaxy as discussed in the previous paragraph, the mass difference of \hi in the inner CGM between the galaxies might imply a corresponding difference in \hi supply to the disk. This could be consistent with the ``bathtub" model of star formation, in which main factor controlling the SFR of a galaxy is the rate of accretion of \hi onto the galaxy disk \citep{Dekel2014}. 

\subsection{Mass}\label{sec:mass}

\noindent We interpret the excess \hi observed by the GBT compared to the convolved WSRT as the \hi in the CGM (see \S\ref{sec:implication}). In this section, we discuss the contribution of this \hi to the total mass budget of neutral CGM, along the minor axes and to the whole CGM of our targets. 

We define the overall halo mass fraction of the \hin, $f_h$ as follows:
    \begin{equation}
    f_h = \frac{M(\hbox{\hin})_{CGM}}{M(\hbox{\hin})_{disk}+M(\hbox{\hin})_{CGM}} 
    \end{equation}
Here, M(\hin)$_{CGM}$ is the total \hi mass in the CGM, traced by our GBT observation. 
M(\hin)$_{disk}$ is the \hi mass of the disk.
Therefore, M(\hin)$_{disk}$ + M(\hin)$_{CGM}$ is the total \hi mass. As we have probed out to the impact parameter of 90--120 kpc from the galaxies, we consider the CGM within this region only. 

First, we consider the CGM along the minor axis, probed directly by our pointings. 
To calculate M(\hin)$_{CGM}$, we subtract the convolved WSRT intensities from the GBT intensities. Then, we convert the diffuse column density estimates to the mass. For this exercise, we use only the pointings spaced by full beam size (i.e., up/down\,1,2,3). Then we sum the mass from each of these pointings (see Table \ref{tab:obs}). For the pointings ``up1/down1", we use the rough estimates obtained in \S\ref{sec:implication}. 
By comparing M(\hin)$_{CGM}$ with M(\hin)$_{disk}$ of NGC\,891 \citep{Oosterloo2007} and NGC\,4565 \citep{Zschaechner2012}, we obtain $f_h$ = $0.014\pm0.001\pm0.002$ and ${0.005}\pm0.001\pm0.001$ for NGC\,891 and NGC\,4565, respectively. This $f_h$ is a lower limit, as this is measured only along the minor axis, not in the whole CGM.  It accounts for about one percent of the total \hi mass, which illustrates why this emission has been so hard to detect. 

We estimate the total mass of diffuse, extended \hi in the CGM assuming an azimuthally symmetric diffuse \hi distribution. At some impact parameters, measurements are available in both sides of the disk (e.g., up/down 1). We consider the average of the intensities at these pointings. Finally, we obtain 
$f_h$ = $0.052\pm0.003^{+0.008}_{-0.010}$ and {$0.020\pm0.005\pm0.006$} for NGC\,891 and NGC\,4565, respectively. As a sanity check, we compare these estimates to those implied by integrating the N(\hin) profiles that we fit in \S\ref{sec:absorption}. Integrating those profiles yields $f_h$ = $0.047\pm0.014$ and {$0.024\pm0.007$} for NGC\,891 and NGC\,4565, in excellent agreement with the values obtained directly from the measured column densities. Thus, adopting simple assumptions, the neutral CGM could, in total, account for roughly two to five percent of the total \hi mass. 

From the area within 50 kpc around NGC\,891 and NGC\,4565, \cite{Pingel2018} estimated that $f_{19}$, the fraction of \hi mass below N(\hin) = 10$^{19}$ cm$^{-2}$ is 0.004--0.006 and 0.007--0.009, respectively. Our measurement is larger than the estimation of \cite{Pingel2018} by almost an order of magnitude. We suggest several potential causes for this discrepancy: (1) The simplest explanation would be that there is a large mass of diffuse, extended \hi that has column density $< 10^{19}$ cm$^{-2}$ and that extends beyond 50\,kpc. This emission might have been missed due to the lower sensitivity of the \citet{Pingel2018} observations. (2) We assumed an azimuthally symmetric diffuse \hi distribution. If our observed \hi is from a feedback-driven outflow or a large-scale galactic fountain, the mass could be preferentially concentrated along the minor axis. In that case, assumption of azimuthal symmetry would lead to an overestimated \hi mass. The lack of offset between the GBT and WSRT data in \cite{Pingel2018} might be due to the extended \hi disk contamination in the azimuthal averages. The diffuse extended \hi is likely abundant away from the major axis. As we focus only on the minor axis, we cannot confirm this in the whole CGM. Deep observations towards other points in the halo, including off-axis directions, will help clarify the situation. 
In any case, $f_h$ differs from $f_{19}$ by definition, so it is not totally clear that whether the results are in conflict or just defined differently.
\section{Conclusion and future directions}\label{sec:conclude}
\noindent We present deep GBT observations along the minor axes of two edge-on spiral galaxies NGC\,891 and NGC\,4565. Below, we summarize our results:
\begin{enumerate}
    \item We have detected \hi out to 90--120 kpc from the center of the galaxies at 5$\sigma$ significance (including systematic uncertainties). These detections imply a column density of $4\times10^{17}$ cm$^{-2}$ and show velocity width of 500--600 km s$^{-1}$ velocity range.
    \item Comparing our GBT measurement to interferometric observations from the WSRT convolved at the same angular resolution of GBT, we find that the \hi intensity in the CGM of NGC\,891 and NGC\,4565 measured by our GBT observations exceeds the WSRT intensity by a factor of $\approx$2. This suggests that we observe a diffuse, extended phase not recovered by the interferometer observations.
    \item The diffuse extended \hi detected along the minor axes are $1.4\pm{0.3}$\% and {$0.5\pm0.1$}\% of the total \hi mass for NGC\,891 and NGC\,4565, including their disks. If we extrapolate these assuming azimuthal symmetry, these escalate to $5.2\pm0.9$\% and {$2.0\pm0.8$}\%, respectively. 
    \item The sign of the average velocities are consistent with an outflow or inflow in NGC\,4565, and a combination of fountain, inflow and outflow in NGC\,891. The absolute values of the average velocities are much smaller than the velocity width, indicating the presence of a quasi-static medium permeating the halo. 
\end{enumerate}  

\noindent To estimate the overall spatial distribution and the mass fraction of the large-scale diffuse \hi in the CGM, it is necessary to extend this single-dish study at larger impact parameter and away from the minor axes of the galaxies. Upcoming interferometric surveys like Widefield ASKAP L-band Legacy All-sky Blind Survey \citep[WALLABY,][]{Kleiner2019} and suggested future surveys planned with SKA (Square Kilometer Array) will significantly increase the probability of detecting small-scale \hi in the CGM. A larger sample of galaxies spanning a range of galactic properties would broaden our understanding of the role diffuse \hi plays in galaxy evolution.
\section*{acknowledgement}
\noindent The Green Bank Observatory is a facility of the National Science Foundation operated under cooperative agreement by Associated Universities, Inc. The work of A.K.L. and M.J.G. was partially supported by the National Science Foundation under Grants No. 1615105, 1615109, and 1653300. AKL is partially supported by NASA ADAP grants NNX16AF48G and NNX17AF39G. A.S. is supported by an NSF Astronomy and Astrophysics Postdoctoral Fellowship under award AST-1903834. DJP acknowledges partial support from NSF CAREER grant AST-114949. 

\facilities{\textit{GBT}, \textit{WSRT}}
\software{GBTIDL v2.10.1 \citep{Marganian2006}, MIRIAD \citep{Sault1995}, NumPy v1.11.3 \citep{Dubois1996}, AstroPy v2.0.6 \citep{Astropy2013}, SciPy v0.17.0 \citep{Oliphant2007}, Matplotlib v1.5.3 \citep{Hunter2007}, DS9 \citep{Joye2003}}

\begin{thebibliography}{}
\expandafter\ifx\csname natexlab\endcsname\relax\def\natexlab#1{#1}\fi
\providecommand{\url}[1]{\href{#1}{#1}}

\bibitem[{{Astropy Collaboration} {et~al.}(2013){Astropy Collaboration},
  {Robitaille}, {Tollerud}, {Greenfield}, {Droettboom}, {Bray}, {Aldcroft},
  {Davis}, {Ginsburg}, \& {Price-Whelan}}]{Astropy2013}
{Astropy Collaboration}, {Robitaille}, T.~P., {Tollerud}, E.~J., {et~al.} 2013,
  \aap, 558, A33

\bibitem[{{Bland-Hawthorn} {et~al.}(2017){Bland-Hawthorn}, {Maloney},
  {Stephens}, {Zovaro}, \& {Popping}}]{Bland-Hawthorn2017}
{Bland-Hawthorn}, J., {Maloney}, P.~R., {Stephens}, A., {Zovaro}, A., \&
  {Popping}, A. 2017, \apj, 849, 51

\bibitem[{{Braun} \& {Thilker}(2004)}]{Braun2004}
{Braun}, R., \& {Thilker}, D.~A. 2004, \aap, 417, 421

\bibitem[{{de Blok} {et~al.}(2014){de Blok}, {Keating}, {Pisano}, {Fraternali},
  {Walter}, {Oosterloo}, {Brinks}, {Bigiel}, \& {Leroy}}]{deBlok2014}
{de Blok}, W.~J.~G., {Keating}, K.~M., {Pisano}, D.~J., {et~al.} 2014, \aap,
  569, A68

\bibitem[{{Dekel} \& {Mandelker}(2014)}]{Dekel2014}
{Dekel}, A., \& {Mandelker}, N. 2014, \mnras, 444, 2071

\bibitem[{{Draine}(2011)}]{Draine2011}
{Draine}, B.~T. 2011, {Physics of the Interstellar and Intergalactic Medium}

\bibitem[{{Dubois} {et~al.}(1996){Dubois}, {Hinsen}, \& {Hugunin}}]{Dubois1996}
{Dubois}, P.~F., {Hinsen}, K., \& {Hugunin}, J. 1996, Computers in Physics, 10,
  262

\bibitem[{{Fraternali}(2017)}]{Fraternali2017}
{Fraternali}, F. 2017, in \apss, Vol. 430, Gas Accretion onto Galaxies, ed.
  A.~{Fox} \& R.~{Dav{\'e}}, 323

\bibitem[{{Fraternali} \& {Binney}(2008)}]{Fraternali2008}
{Fraternali}, F., \& {Binney}, J.~J. 2008, \mnras, 386, 935

\bibitem[{{Heald} {et~al.}(2011){Heald}, {J{\'o}zsa}, {Serra}, {Zschaechner},
  {Rand}, {Fraternali}, {Oosterloo}, {Walterbos}, {J{\"u}tte}, \&
  {Gentile}}]{Heald2011}
{Heald}, G., {J{\'o}zsa}, G., {Serra}, P., {et~al.} 2011, \aap, 526, A118

\bibitem[{{Hibbard} \& {van Gorkom}(1996)}]{Hibbard1996}
{Hibbard}, J.~E., \& {van Gorkom}, J.~H. 1996, \aj, 111, 655

\bibitem[{{Hodges-Kluck} {et~al.}(2018){Hodges-Kluck}, {Bregman}, \&
  {Li}}]{Hodges-Kluck2018}
{Hodges-Kluck}, E.~J., {Bregman}, J.~N., \& {Li}, J.-t. 2018, \apj, 866, 126

\bibitem[{{Howk} \& {Savage}(1997)}]{Howk1997}
{Howk}, J.~C., \& {Savage}, B.~D. 1997, \aj, 114, 2463

\bibitem[{{Howk} \& {Savage}(1999)}]{Howk1999}
---. 1999, \aj, 117, 2077

\bibitem[{{Hunter}(2007)}]{Hunter2007}
{Hunter}, J.~D. 2007, Computing in Science and Engineering, 9, 90

\bibitem[{{Joung} {et~al.}(2012){Joung}, {Bryan}, \& {Putman}}]{Joung2012}
{Joung}, M.~R., {Bryan}, G.~L., \& {Putman}, M.~E. 2012, \apj, 745, 148

\bibitem[{{Joye} \& {Mandel}(2003)}]{Joye2003}
{Joye}, W.~A., \& {Mandel}, E. 2003, in Astronomical Society of the Pacific
  Conference Series, Vol. 295, Astronomical Data Analysis Software and Systems
  XII, ed. H.~E. {Payne}, R.~I. {Jedrzejewski}, \& R.~N. {Hook}, 489

\bibitem[{{Kere{\v{s}}} {et~al.}(2005){Kere{\v{s}}}, {Katz}, {Weinberg}, \&
  {Dav{\'e}}}]{Keres2005}
{Kere{\v{s}}}, D., {Katz}, N., {Weinberg}, D.~H., \& {Dav{\'e}}, R. 2005,
  \mnras, 363, 2

\bibitem[{{Kleiner} {et~al.}(2019){Kleiner}, {Koribalski}, {Serra}, {Whiting},
  {Westmeier}, {Wong}, {Kamphuis}, {Popping}, {Bekiaris}, {Elagali}, {For},
  {Lee-Waddell}, {Madrid}, {Reynolds}, {Rhee}, {Shao}, {Staveley-Smith},
  {Wang}, {Anderson}, {Collier}, {Ord}, \& {Voronkov}}]{Kleiner2019}
{Kleiner}, D., {Koribalski}, B.~S., {Serra}, P., {et~al.} 2019, \mnras, 488,
  5352

\bibitem[{{Koribalski} {et~al.}(2018){Koribalski}, {Wang}, {Kamphuis},
  {Westmeier}, {Staveley-Smith}, {Oh}, {L{\'o}pez-S{\'a}nchez}, {Wong}, {Ott},
  {de Blok}, \& {Shao}}]{Koribalski2018}
{Koribalski}, B.~S., {Wang}, J., {Kamphuis}, P., {et~al.} 2018, \mnras, 478,
  1611

\bibitem[{{Lockman} {et~al.}(2012){Lockman}, {Free}, \&
  {Shields}}]{Lockman2012}
{Lockman}, F.~J., {Free}, N.~L., \& {Shields}, J.~C. 2012, \aj, 144, 52

\bibitem[{{Marganian} {et~al.}(2006){Marganian}, {Garwood}, {Braatz},
  {Radziwill}, \& {Maddalena}}]{Marganian2006}
{Marganian}, P., {Garwood}, R.~W., {Braatz}, J.~A., {Radziwill}, N.~M., \&
  {Maddalena}, R.~J. 2006, in Astronomical Society of the Pacific Conference
  Series, Vol. 351, Astronomical Data Analysis Software and Systems XV, ed.
  C.~{Gabriel}, C.~{Arviset}, D.~{Ponz}, \& S.~{Enrique}, 512

\bibitem[{{Mould} \& {Sakai}(2008)}]{Mould2008}
{Mould}, J., \& {Sakai}, S. 2008, \apjl, 686, L75

\bibitem[{{Nelson} {et~al.}(2013){Nelson}, {Vogelsberger}, {Genel}, {Sijacki},
  {Kere{\v{s}}}, {Springel}, \& {Hernquist}}]{Nelson2013}
{Nelson}, D., {Vogelsberger}, M., {Genel}, S., {et~al.} 2013, \mnras, 429, 3353

\bibitem[{{Odekon} {et~al.}(2016){Odekon}, {Koopmann}, {Haynes}, {Finn},
  {McGowan}, {Micula}, {Reed}, {Giovanelli}, \& {Hallenbeck}}]{Odekon2016}
{Odekon}, M.~C., {Koopmann}, R.~A., {Haynes}, M.~P., {et~al.} 2016, \apj, 824,
  110

\bibitem[{{Oliphant}(2007)}]{Oliphant2007}
{Oliphant}, T.~E. 2007, Computing in Science and Engineering, 9, 10

\bibitem[{{Oosterloo} {et~al.}(2007){Oosterloo}, {Fraternali}, \&
  {Sancisi}}]{Oosterloo2007}
{Oosterloo}, T., {Fraternali}, F., \& {Sancisi}, R. 2007, \aj, 134, 1019

\bibitem[{{Pingel} {et~al.}(2018){Pingel}, {Pisano}, {Heald}, {Jarrett}, {de
  Blok}, {J{\'o}zsa}, {J{\"u}tte}, {Rand}, {Oosterloo}, \&
  {Winkel}}]{Pingel2018}
{Pingel}, N.~M., {Pisano}, D.~J., {Heald}, G., {et~al.} 2018, \apj, 865, 36

\bibitem[{{Pisano}(2014)}]{Pisano2014}
{Pisano}, D.~J. 2014, \aj, 147, 48

\bibitem[{{Popping} {et~al.}(2009){Popping}, {Dav{\'e}}, {Braun}, \&
  {Oppenheimer}}]{Popping2009}
{Popping}, A., {Dav{\'e}}, R., {Braun}, R., \& {Oppenheimer}, B.~D. 2009, \aap,
  504, 15

\bibitem[{{Prochaska} {et~al.}(2017){Prochaska}, {Werk}, {Worseck}, {Tripp},
  {Tumlinson}, {Burchett}, {Fox}, {Fumagalli}, {Lehner}, \&
  {Peeples}}]{Prochaska2017}
{Prochaska}, J.~X., {Werk}, J.~K., {Worseck}, G., {et~al.} 2017, \apj, 837, 169

\bibitem[{{Putman} {et~al.}(2012){Putman}, {Peek}, \& {Joung}}]{Putman2012}
{Putman}, M.~E., {Peek}, J.~E.~G., \& {Joung}, M.~R. 2012, \araa, 50, 491

\bibitem[{{Qu} {et~al.}(2019){Qu}, {Bregman}, \& {Hodges-Kluck}}]{Qu2019}
{Qu}, Z., {Bregman}, J.~N., \& {Hodges-Kluck}, E.~J. 2019, \apj, 876, 101

\bibitem[{{Radburn-Smith} {et~al.}(2011){Radburn-Smith}, {de Jong}, {Seth},
  {Bailin}, {Bell}, {Brown}, {Bullock}, {Courteau}, {Dalcanton}, {Ferguson},
  {Goudfrooij}, {Holfeltz}, {Holwerda}, {Purcell}, {Sick}, {Streich}, {Vlajic},
  \& {Zucker}}]{Radburn-Smith2011}
{Radburn-Smith}, D.~J., {de Jong}, R.~S., {Seth}, A.~C., {et~al.} 2011, \apjs,
  195, 18

\bibitem[{{Rand}(1998)}]{Rand1998}
{Rand}, R.~J. 1998, \apj, 501, 137

\bibitem[{{Rand} {et~al.}(1992){Rand}, {Kulkarni}, \& {Hester}}]{Rand1992}
{Rand}, R.~J., {Kulkarni}, S.~R., \& {Hester}, J.~J. 1992, \apj, 396, 97

\bibitem[{{Rees} \& {Ostriker}(1977)}]{Rees1977}
{Rees}, M.~J., \& {Ostriker}, J.~P. 1977, \mnras, 179, 541

\bibitem[{{Ribaudo} {et~al.}(2011){Ribaudo}, {Lehner}, {Howk}, {Werk}, {Tripp},
  {Prochaska}, {Meiring}, \& {Tumlinson}}]{Ribaudo2011}
{Ribaudo}, J., {Lehner}, N., {Howk}, J.~C., {et~al.} 2011, \apj, 743, 207

\bibitem[{{Richter} {et~al.}(2017){Richter}, {Nuza}, {Fox}, {Wakker}, {Lehner},
  {Ben Bekhti}, {Fechner}, {Wendt}, {Howk}, {Muzahid}, {Ganguly}, \&
  {Charlton}}]{Richter2017}
{Richter}, P., {Nuza}, S.~E., {Fox}, A.~J., {et~al.} 2017, \aap, 607, A48

\bibitem[{{Rupen}(1991)}]{Rupen1991}
{Rupen}, M.~P. 1991, \aj, 102, 48

\bibitem[{{Sancisi} {et~al.}(2008){Sancisi}, {Fraternali}, {Oosterloo}, \& {van
  der Hulst}}]{Sancisi2008}
{Sancisi}, R., {Fraternali}, F., {Oosterloo}, T., \& {van der Hulst}, T. 2008,
  \aapr, 15, 189

\bibitem[{{Sault} {et~al.}(1995){Sault}, {Teuben}, \& {Wright}}]{Sault1995}
{Sault}, R.~J., {Teuben}, P.~J., \& {Wright}, M.~C.~H. 1995, in Astronomical
  Society of the Pacific Conference Series, Vol.~77, Astronomical Data Analysis
  Software and Systems IV, ed. R.~A. {Shaw}, H.~E. {Payne}, \& J.~J.~E.
  {Hayes}, 433

\bibitem[{{Schmidt} {et~al.}(2019){Schmidt}, {Krause}, {Heesen}, {Basu},
  {Beck}, {Wiegert}, {Irwin}, {Heald}, {Rand}, {Li}, \& {Murphy}}]{Schmidt2019}
{Schmidt}, P., {Krause}, M., {Heesen}, V., {et~al.} 2019, \aap, 632, A12

\bibitem[{{Sorgho} {et~al.}(2019){Sorgho}, {Carignan}, {Pisano}, {Oosterloo},
  {de Blok}, {Korsaga}, {Pingel}, {Sardone}, {Goedhart}, {Passmoor}, {Dikgale},
  \& {Sirothia}}]{Sorgho2019}
{Sorgho}, A., {Carignan}, C., {Pisano}, D.~J., {et~al.} 2019, \mnras, 482, 1248

\bibitem[{{Stocke} {et~al.}(2010){Stocke}, {Keeney}, \&
  {Danforth}}]{Stocke2010}
{Stocke}, J.~T., {Keeney}, B.~A., \& {Danforth}, C.~W. 2010, \pasa, 27, 256

\bibitem[{{Strickland} {et~al.}(2004){Strickland}, {Heckman}, {Colbert},
  {Hoopes}, \& {Weaver}}]{Strickland2004a}
{Strickland}, D.~K., {Heckman}, T.~M., {Colbert}, E. J.~M., {Hoopes}, C.~G., \&
  {Weaver}, K.~A. 2004, \apjs, 151, 193

\bibitem[{{Tilton} {et~al.}(2012){Tilton}, {Danforth}, {Shull}, \&
  {Ross}}]{Tilton2012}
{Tilton}, E.~M., {Danforth}, C.~W., {Shull}, J.~M., \& {Ross}, T.~L. 2012,
  \apj, 759, 112

\bibitem[{{Tripp} {et~al.}(2008){Tripp}, {Sembach}, {Bowen}, {Savage},
  {Jenkins}, {Lehner}, \& {Richter}}]{Tripp2008}
{Tripp}, T.~M., {Sembach}, K.~R., {Bowen}, D.~V., {et~al.} 2008, \apjs, 177, 39

\bibitem[{{T{\"u}llmann} {et~al.}(2006){T{\"u}llmann}, {Pietsch}, {Rossa},
  {Breitschwerdt}, \& {Dettmar}}]{Tullmann2006}
{T{\"u}llmann}, R., {Pietsch}, W., {Rossa}, J., {Breitschwerdt}, D., \&
  {Dettmar}, R.~J. 2006, \aap, 448, 43

\bibitem[{{Tumlinson} {et~al.}(2017){Tumlinson}, {Peeples}, \&
  {Werk}}]{Tumlinson2017}
{Tumlinson}, J., {Peeples}, M.~S., \& {Werk}, J.~K. 2017, \araa, 55, 389

\bibitem[{{Tumlinson} {et~al.}(2013){Tumlinson}, {Thom}, {Werk}, {Prochaska},
  {Tripp}, {Katz}, {Dav{\'e}}, {Oppenheimer}, {Meiring}, {Ford}, {O'Meara},
  {Peeples}, {Sembach}, \& {Weinberg}}]{Tumlinson2013}
{Tumlinson}, J., {Thom}, C., {Werk}, J.~K., {et~al.} 2013, \apj, 777, 59

\bibitem[{{Veilleux} {et~al.}(2020){Veilleux}, {Maiolino}, {Bolatto}, \&
  {Aalto}}]{Veilleux2020}
{Veilleux}, S., {Maiolino}, R., {Bolatto}, A.~D., \& {Aalto}, S. 2020, \aapr,
  28, 2

\bibitem[{{Veron-Cetty} \& {Veron}(2000)}]{Veroncetty2000}
{Veron-Cetty}, M.~P., \& {Veron}, P. 2000, {A catalogue of quasars and active
  nuclei}

\bibitem[{{Voit} {et~al.}(2017){Voit}, {Meece}, {Li}, {O'Shea}, {Bryan}, \&
  {Donahue}}]{Voit2017}
{Voit}, G.~M., {Meece}, G., {Li}, Y., {et~al.} 2017, \apj, 845, 80

\bibitem[{{Wolfe} {et~al.}(2013){Wolfe}, {Pisano}, {Lockman}, {McGaugh}, \&
  {Shaya}}]{Wolfe2013}
{Wolfe}, S.~A., {Pisano}, D.~J., {Lockman}, F.~J., {McGaugh}, S.~S., \&
  {Shaya}, E.~J. 2013, \nat, 497, 224

\bibitem[{{Wotta} {et~al.}(2016){Wotta}, {Lehner}, {Howk}, {O'Meara}, \&
  {Prochaska}}]{Wotta2016}
{Wotta}, C.~B., {Lehner}, N., {Howk}, J.~C., {O'Meara}, J.~M., \& {Prochaska},
  J.~X. 2016, \apj, 831, 95

\bibitem[{{Zschaechner} {et~al.}(2012){Zschaechner}, {Rand}, {Heald},
  {Gentile}, \& {J{\'o}zsa}}]{Zschaechner2012}
{Zschaechner}, L.~K., {Rand}, R.~J., {Heald}, G.~H., {Gentile}, G., \&
  {J{\'o}zsa}, G. 2012, \apj, 760, 37

\end{thebibliography}
\bibliographystyle{aasjournal}



\end{document}